\documentclass[journal=jacsat,manuscript=article,traditional]{achemso}

\usepackage{chemformula} 
\usepackage[T1]{fontenc} 

\usepackage[latin1]{inputenc}
\usepackage{amsmath}
\usepackage{amsfonts}
\usepackage{hyperref}
\usepackage{amssymb}
\usepackage{graphicx}
\usepackage{hyperref}
\usepackage{colortbl}
\usepackage{subfigure}
\usepackage{gensymb}
\usepackage{svg}
\usepackage{verbatim}
\usepackage{soul}

\mciteErrorOnUnknownfalse

\usepackage[T1]{fontenc}
\usepackage{mathptmx}
\usepackage{bbold}
\usepackage{scalerel}
\usepackage{xcolor}
\usepackage{hyperref}
\usepackage{mathtools}

\renewcommand*\rm[1]{\ensuremath{\mathrm{#1}}}
\renewcommand*\bf[1]{\ensuremath{\mathbf{#1}}}
\newcommand{\hyref}[1]{\hyperref[#1]{\ref{#1}}}
\newcommand{\dd}{\mathrm{d}}

\newcommand{\beginsupplement}{
        \setcounter{table}{0}
        \renewcommand{\thetable}{S\arabic{table}}
        \setcounter{figure}{0}
        \renewcommand{\thefigure}{S\arabic{figure}}%
     }

\author{M. Aarts}
\affiliation[AMOLF]
{Center for Nanophotonics, AMOLF, Science Park 109, Amsterdam, Netherlands}
\altaffiliation{Authors contributed equally}
\author{W.Q.Boon}
\altaffiliation{Authors contributed equally}
\affiliation[UU1]
{Institute for Theoretical Physics, Utrecht University , Princetonplein 5, 3584 CC Utrecht}
\altaffiliation{Authors contributed equally}
\author{B. Cu\'enod}
\affiliation[AMOLF]
{Center for Nanophotonics, AMOLF, Science Park 109, Amsterdam, Netherlands}
\author{M. Dijkstra}
\affiliation[UU2]
{Soft Condensed Matter, Debye Institute for Nanomaterials Science, Utrecht University, Princetonplein 1, 3584 CC Utrecht, The Netherlands}
\author{R. van Roij}
\affiliation[UU1]
{Institute for Theoretical Physics, Utrecht University , Princetonplein 5, 3584 CC Utrecht}
\author{E. Alarcon-Llado}
\affiliation[AMOLF]
{Center for Nanophotonics, AMOLF, Science Park 109, Amsterdam, Netherlands}
\email{e.alarconllado@amolf.nl}

\title{Ion Current Rectification and Long-Range Interference in Conical Silicon Micropores }

\abbreviations{}

\begin{document}

\begin{abstract}
Fluidic devices exhibiting ion current rectification (ICR), or ionic diodes, are of broad interest for applications including desalination, energy harvesting, and sensing, amongst others. For such applications a large conductance is desirable which can be achieved by simultaneously using thin membranes and wide pores. In this paper we demonstrate ICR in micron sized conical channels in a thin silicon membrane with pore diameters comparable to the membrane thickness but both much larger than the electrolyte screening length. We show that for these pores the entrance resistance is not only key to Ohmic conductance around 0 V, but also for understanding ICR, both of which we measure experimentally and capture within a single analytic theoretical framework. The only fit parameter in this theory is the membrane surface potential, for which we find that it is voltage dependent and its value is excessively large compared to literature. From this we infer that surface charge outside the pore strongly contributes to the observed Ohmic conductance and rectification by a different extent. We experimentally verify this hypothesis in a small array of pores and find that ICR vanishes due to pore-pore interactions mediated through the membrane surface, while Ohmic conductance around 0 V remains unaffected. We find that the pore-pore interaction for ICR is set by a long-ranged decay of the concentration which explains the surprising finding that the ICR vanishes for even a sparsely populated array with a pore-pore spacing as large as 7 $\mu$m. 
\end{abstract}

\begin{figure}[h!]
	\includegraphics[width=0.5\textwidth]{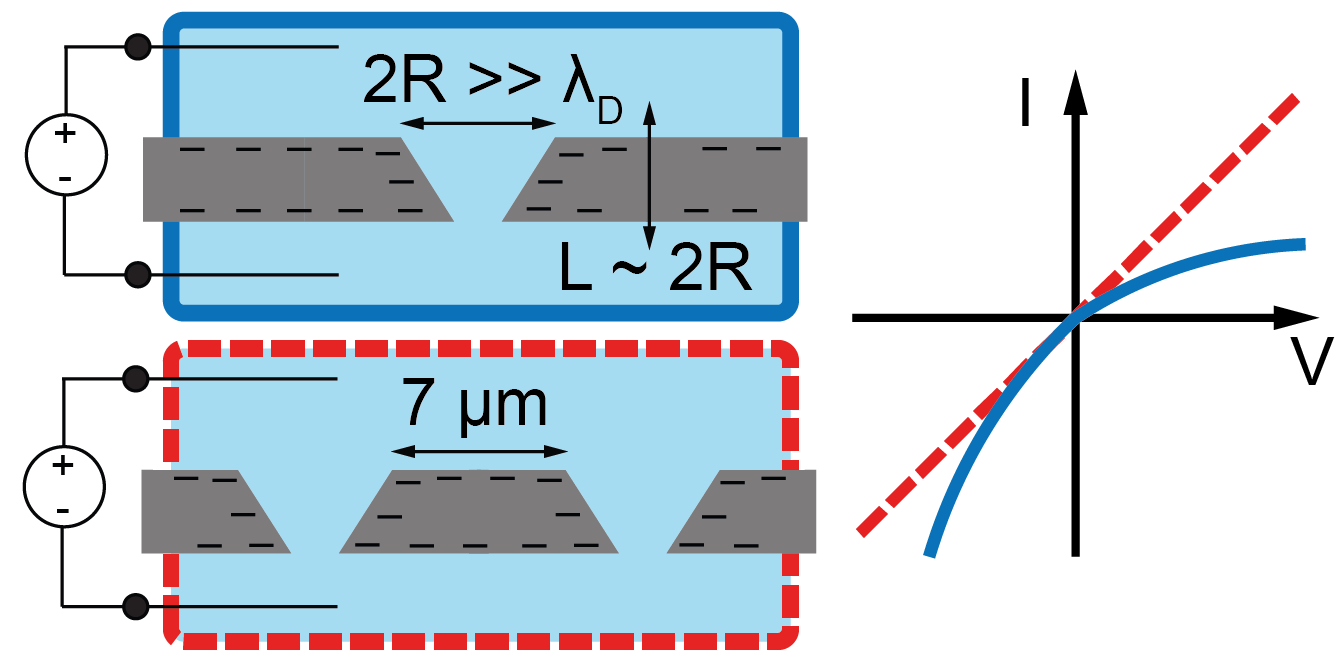}
	\label{TOC}
\end{figure}

\section{Introduction}
Ionic transport near solid-liquid interfaces can differ drastically from that in  bulk due to Coulombic interactions with the surface \cite{Faucher2019}. Such interface effects can be used to tailor nanofluidic devices \cite{Bocquet2020}, finding applications in desalination \cite{Nishizawa1995,Vlassiouk2008}, ionic circuitry \cite{Daiguji2005,Kalman2008}, bio-chemical sensing\cite{biochem1,biochem2,biochem3,biochem4, biochem5}, energy harvesting \cite{VanderHeyden2006,Xiao2019} and neuromorphic signalling\cite{wang_hysteresis, wang_hysteresis2, iontronics_review}. 
A particularly useful element for directional control of ionic currents is a current rectifier \cite{Siria2017, hou_diode, bush_diode, biochem3}, also known as a diode. In fact, the phenomenon of ion current rectification (ICR) has been observed and extensively studied in nanochannels \cite{Xiong2019,woermann1, woermann2, conereview}. \\
The ICR originates from an asymmetry in the ionic current along the length of the channel, due to a varying relative contribution to the ionic current of the charge-selective electric double layer (EDL) that screens the charge on the channel walls. Typically, ICR is demonstrated in nanoscale conical channels, where EDL overlap occurs on the narrow end of the channel\cite{woermann1, woermann2}. In general, the ICR mechanism for a geometrically asymmetric, or tapered, channel can be understood by considering that the relative contribution of the salt current through the EDL to the total current is smaller at the wide opening than at the narrow opening \cite{Siwy2006, POF}. This results in an asymmetry of the transference (i.e. the partial current due to either ionic species). Considering a channel with a negative surface charge on its wall, resulting in an EDL with excess positive ionic charge, an electric field directed towards the narrow end leads to more  ions leaving the small opening than entering the large opening (before steady state is reached), resulting in depletion of ions inside the channel, and a suppressed conductance \cite{Poggioli2019}. The opposite is true for an oppositely directed electric field, resulting in accumulation of charge carriers and enhanced conductance. More broadly, the required asymmetry in transference can be introduced not only by geometry, but also by a variation of e.g. charge or concentration. \cite{Siwy2006,Cheng2007,Poggioli2019} \\
For application purposes regarding larger scale porous membranes, a low electric resistance of the channel is desirable to mitigate Ohmic losses. Two intuitive ways to construct a channel with low resistance are by making (i) larger openings\cite{Poggioli2019} or (ii) shorter channels \cite{Graf2019, siwy_saturation}. Considering the accumulation/depletion mechanism described above, recent theoretical work predicts that ICR can also occur in wide channels without overlapping electric double layers, as long as a substantial part of the ionic current is due to surface conductance \cite{Poggioli2019,Cengio2019, rect1largeR, Lin2018, cervera_exp_micropores_and_saturation, zhou_exp_micropores}. In fact, ICR in mesoscopic channels \cite{Jubin2018} and chemically modified micronsized systems have recently been observed \cite{Lin2018,Lin2019,He2017}. For thin membranes with short channels on the other hand, it has become clear that the applied potential partially drops outside the channel, rather than fully over the channel itself \cite{Lee2012, hall1975access, siwy_saturation, pyramidal, yossifon}. These extended entrance effects give rise to an edge, or access, resistance and become relevant for the behavior of a fluidic pore with a channel length of the order of the diameter, which can either positively contribute to ICR or interfere destructively in arrays of pores \cite{Ma2018,Su2018, yossifon}. As of now, however, such pore-pore interactions are still poorly understood.  \\
In this work we fabricate conical, i.e. geometrically asymmetric, fluidic micropores in thin (2 $\mu$m) crystalline silicon membranes, with base and tip radii of $R_{\rm{b}}\approx$ 1.5 $ \mu$m and $R_{\rm{t}}\approx$ 0.5 $ \mu$m, respectively, such that even the smallest of these channel dimensions is larger than the typical electrolyte screening length by more than an order of magnitude. We demonstrate that these pores exhibit ion current rectification, and we develop an analytical theory for the channel conductance in which the surface potential is the only fit parameter. We stress that the (Ohmic) channel conductance at low applied potentials and ICR are distinct phenomena and we find that we need a different surface potential to fit the experimental data to these two effects, with both surface potentials being very large, implying a very large surface charge. We interpret the value of these fitted surface potentials as non-physical, and rather attribute this excessively large charge to a contribution of conduction along the planar membrane surface outside the channel at the inlet and outlet of the pore unaccounted for in our model. By correcting this required surface charge to an effective area, we estimate that this membrane surface conduction is relevant up to distances around the pore opening as large as 7.4 $\mu$m for Ohmic channel conductance, and 15.0 $\mu$m for ICR, implying that a larger area around the pore is required for ICR. We test this hypothesis by fabricating a small array of pores with a 10 $\mu$m spacing ($\approx$ 10$^{6}$ pores$/$cm$^{2}$). Indeed, we find that despite this low pore density the Ohmic conductance remains unaffected, but that the ICR vanishes for the array. Extended entrance effects at the micron scale therefore appear to play a significant role in the required asymmetry in ion transport through pores in thin membranes, which we attribute to the long-ranged decay of the electric field outside the pore. This electric field creates a concentration profile with a similar long-ranged, inverse-square with distance, decay into the bulk. This scale-free decay introduces long-ranged pore-pore interactions for thin pores, which become particularly relevant in array configurations typical for membranes. 

\section{Experimental}

\begin{figure}[b!]
	\includegraphics[width=0.7\textwidth]{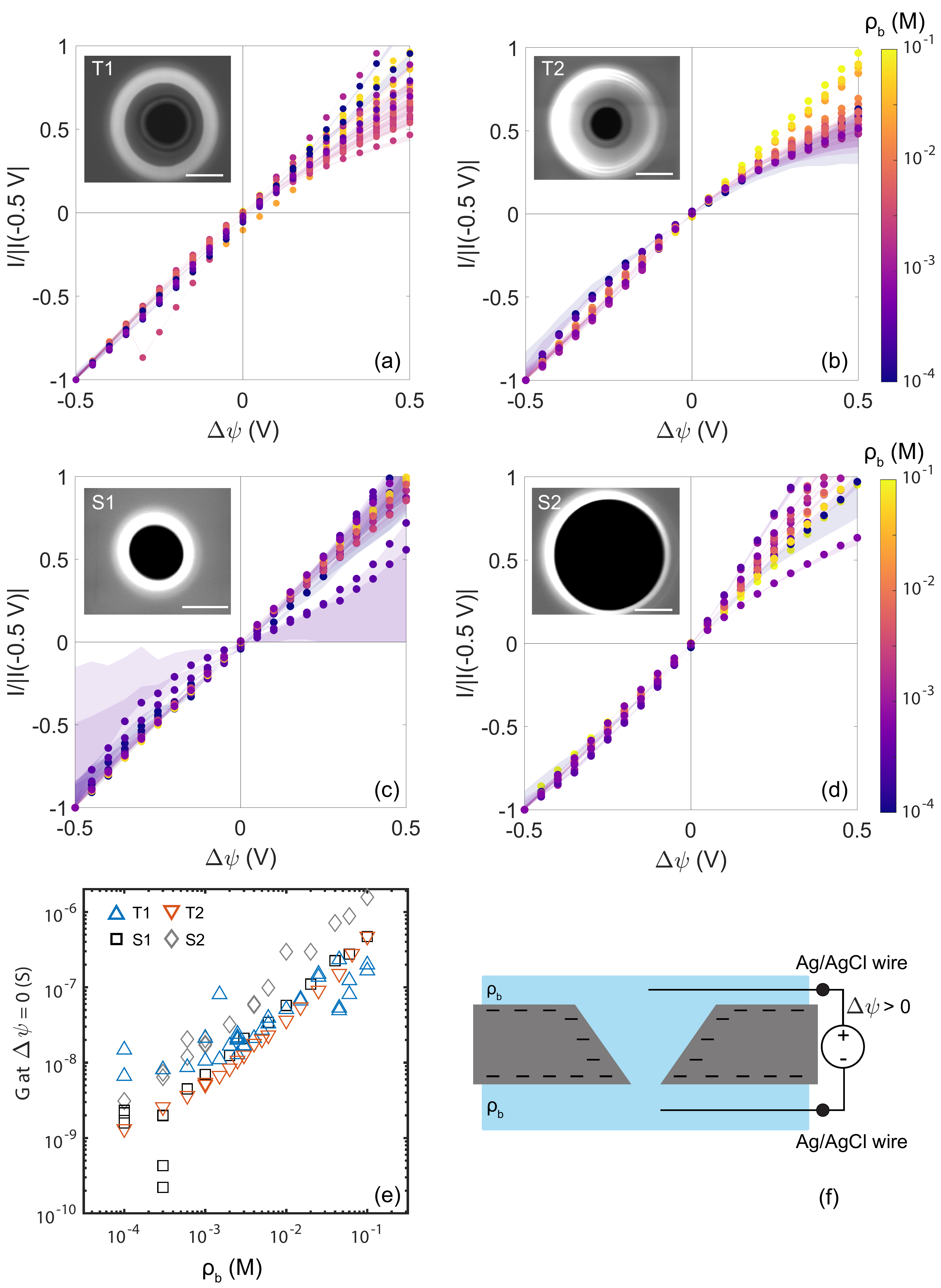}
	\caption{(a-d) Experimental current-voltage (I-V) curves, normalized at a potential drop of $-0.5$ V at various KCl concentrations $\rho_{\rm{b}}\in[10^{-4}-10^{-1}]$ M indicated by the colour scale. The shaded region indicate uncertainty in the measurement due to the leakage current obtained from an as-received membrane (SI-3). Systematic current rectification is observed for tapered pores T1 and T2 with the conductivity at $+0.5$ V being lower than at $-0.5$ V. The inset shows scanning electron microscopy images of the tapered pores T1 (a) and T2 (b) and straight pores S1 (c) and S2 (d) after fabrication. The scalebars are 1 $\mu$m. (e) Conductance of the pores at 0 V as a function of KCl concentration $\rho_{\rm{b}}(M)$. (f) Schematic of the experimental setup where 2 aqueous reservoirs of equal KCl concentration are separated by the membrane with a single pore. The polarity of the potential is such that positive potentials indicate the anode being in the reservoir facing the large opening of the pore.}
	\label{Fig2}
\end{figure}

For our conductance measurements we fabricate single micron-sized pores, which are either straight or tapered, in 2 $\pm$ 0.5 $\mu$m thick crystalline silicon membranes using a focused ion beam (FIB). Si is a reliable and cross-compatible platform that allows for precise pore manufacturing. The taper is created by writing concentric circles with decreasing radius, resulting in an asymmetric pore, as verified by atomic force microscopy in SI-1. Conductance measurements are carried out by placing the membrane between two aqueous reservoirs containing KCl of equal bulk concentration ($\rho_{\rm{b}}$), and applying a potential between the reservoirs using Ag/AgCl wire electrodes (Figure \ref{Fig2} (f), and SI-2). Of note is the polarity of the applied potential, where positive potentials indicate the anode being in the reservoir facing the large opening of the pore.\\
We record quasi-static current-voltage (I-V) curves between -0.5 V and 0.5 V (see methods) at different reservoir KCl bulk concentrations $\rho_{\rm{b}}$ ranging from 10$^{-1}$ M to 10$^{-4}$ M, for 4 membranes containing a single pore. The insets of Figure \ref{Fig2}(a)-(d) show scanning electron microscopy (SEM) images directly after fabrication of the tapered pores T1 (base and tip radii $R_{\rm{b}}\simeq$ $1.5\mu$m,$\ R_{\rm{t}}\simeq 0.5$ $\mu$m) and T2 ($R_{\rm{b}}\simeq$ $1.5\mu$m,$\ R_{\rm{t}}\simeq 0.4$ $\mu$m) and two straight reference pores S1 ($R_{\rm{b}}=R_{\rm{t}}\simeq0.6$ $\mu$m) and S2 ($R_{\rm{b}}=R_{\rm{t}}\simeq1.5$ $\mu$m).\\
The corresponding I-V curves are shown as circles in Figure \ref{Fig2}(a)-(d), where the colors label the salt concentration $\rho_{\rm{b}}$. As the magnitude of the current response varies by several orders of magnitude over the salt concentration range, the current is normalized to the value at an applied potential of -0.5 V for visibility. The shaded regions indicate the possible contribution from leakage current through the membrane, averaged from measurements on an as-received membrane without a pore (SI-3). Due to the range in magnitude of the measured currents, this is most relevant for the lowest concentrations and the smallest pore (S1). The conductance at 0 V as a function of concentration is shown in Figure \ref{Fig2}(e). A linear decrease of the conductance with decreasing concentration is observed with the conductance saturating at $\rho_{\rm{b}}$ < 1 mM. \\
At the highest concentrations (yellow, $\rho_{\rm{b}}$ = 0.1 M), and therefore the smallest Debye length ($\lambda_{\rm{D}}$ $\simeq$ 1 nm), all pores show a linear I-V response, consistent with bulk dominated transport. At lower concentrations, however, conductance through the tapered channels starts to show ion current rectification. It should be noted that even at the lowest concentration, $\rho_{\rm{b}}$ = $0.1$ mM, the electrolyte screening length $\lambda_{\rm{D}}$ $\simeq$ 30 nm is much shorter than the smallest tip radius, so that the micropores are well outside the regime of EDL overlap. While some curves for the straight pores S1 and S2 show an erratic deviation from ideal symmetrical conductance, the tapered pores T1 and T2 show systematic modulation of rectification, where the conductance at positive potentials is smaller than that at negative potentials.

\subsection{Theoretical framework}

\begin{figure}[b!]
	\includegraphics[width=1\textwidth]{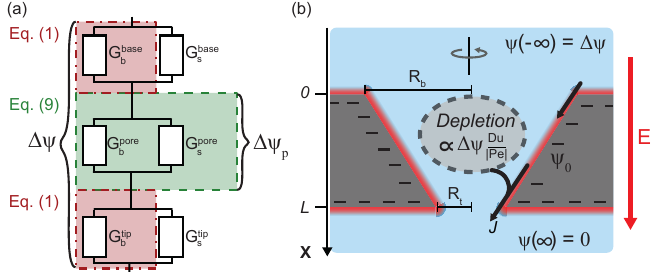}
	\caption{(a) Schematic equivalent circuit of a pore featuring bulk and surface conductance $G_{\rm{b}}^i$ and $G_{\rm{s}}^i$ respectively, at the base-, tip- and within the pore-regrion. The elements considered in our analytical model are highlighted, where part of the applied potential $\Delta\psi$ drops over the edge resistance (red dash-dotted region), as captured by Eq.(\hyref{pot_drop}). The conductance of the pore with the remaining potential $\Delta\psi_{\rm{p}}$ (green dashed region) is described by Eq.(\hyref{gen_conductance}). (b) Representation of the conical system under consideration with base and tip radii $R_{\rm{b}}$ and $R_{\rm{t}}$, respectively, and an electric field $-\partial_x\psi=E$ pointed towards the tip. As outlined in the main text, the depletion of ions in the channel is proportional to the potential drop $\Delta\psi$ times the ratio of the Dukhin (Du) and P\'{e}clet (Pe) numbers as shown in Eq.(\hyref{eq:concentrationa}) and illustrated here for a channel wall with a negative surface charge resulting in depletion for a positive potential drop due to a salt flux $J$ (black arrow) through the electric double layer (EDL) that increases towards the tip.}
	\label{Cartoon_theory}
\end{figure}
In the following, we present a model for the potential-dependent conductance of a tapered pore and obtain a closed-form expression that simultaneously describes Ohmic conductance and ion current rectification. Currently in the literature there are two, complementary, theories for current rectification without EDL overlap for pores with large aspect ratio. The theory by Cengio\cite{Cengio2019} and Poggioli\cite{Poggioli2019} describes ICR through the variation of the surface conductance over the pore length but neglects electro-osmotic flow, while the theory of Ref.\cite{POF} (developed by some of the present authors) does account for this flow but fails at extremely low salt concentrations. Hence both theories are complementary rather than mutually exclusive: Refs.\cite{Cengio2019, Poggioli2019} are valid at all concentrations while theory Ref.\cite{POF} is valid at all flow rates. We will find that our experiments show characteristic flow-sensitive behavior, and therefore we build on the theory of Ref.\cite{POF}. However, to describe our experiments either theory would need to be extended as the membranes thickness here is similar to the radius of the pores, and the theory therefore has to account for the electric edge resistance which is comparable to the pore resistance. There is a variety of theories available in the literature accounting for Ohmic edge resistance\cite{Lee2012, Bocquet2020, pyramidal, siwy_saturation, dekker_resistance, hall1975access},the most commonly used one by Hall\cite{hall1975access} which we reproduce here. In this work we will find that not only do edge resistances alter Ohmic conductance, they also significantly alter rectification. In the following section we extend the theory of Ref.\cite{POF} to account for edge effects.
 
 The introduction of non-negligible edge resistances implies an equivalent electric circuit as illustrated in Figure \ref{Cartoon_theory}(a), which not only considers bulk and surface contributions to the conductance inside the pore ($G^{\rm{pore}}_{\rm{b}}$ and $G^{\rm{pore}}_{\rm{\rm{s}}}$) in parallel\cite{kirby_micro, IUPAC} but also two base- ($G^{\rm{base}}_{\rm{b}}$ and $G^{\rm{base}}_{\rm{\rm{s}}}$) and tip- ($G^{\rm{tip}}_{\rm{b}}$ and $G^{\rm{tip}}_{\rm{\rm{s}}}$) conductances. Only recently it has been shown that the charged surface outside of the membrane contributes to edge resistance\cite{Lee2012,Bocquet2020}, but unfortunately our model is not able to explicitly account for the base- and tip surface conductances ($G^{\rm{base}}_{\rm{\rm{s}}}$ and $G^{\rm{tip}}_{\rm{\rm{s}}}$, respectively) even though we will see these charged regions do contribute significantly to both experimental Ohmic conductance and ICR. Instead we implicitly account for the charge on the outside membrane through an "apparent" (large) surface charge within the pore, inflating the pore surface conductance $G_{\rm{s}}^{\rm{pore}}$ and total conductance $G$. Hence our large "apparent" surface charge will account for outer-membrane conductance increasing the well-known bulk edge conductances as described by Hall\cite{hall1975access}. These parallel edge surface and bulk conductances are in series with the pore resistances as depicted in Fig.\hyref{Cartoon_theory}(a), which for the present system parameters ensures that the potential drop over the pore $\Delta\psi_{\rm{p}}$ is significantly smaller than the total bias $\Delta\psi$. This decreased potential drop does not only reduce the current through the pore but also lowers the electro-osmotic flow and concentration polarization within the pore. To obtain $\Delta\psi_{\rm{p}}$ we consider, in cylindrical $(x,r,\theta)$ coordinates, two bulk reservoirs in the half spaces $x<0$ and $x>L$ connected by an azimuthally-symmetric conical channel of length $L$ as depicted in Figure \ref{Cartoon_theory}(b) with base radius $R_{\rm{b}}$ at the inlet ($x=0$) and tip radius $ R_{\rm{t}}$ at the outlet ($x=L$), such that the radius of the channel reads $R(x)=R_{\rm{b}}- (x/L) (R_{\rm{b}}-R_{\rm{t}})$ for $x\in[0,L]$. The potential drop over the inside of the pore can be calculated using two assumptions: (i) that the electric field at the tip (and base) decays as a monopole $-\nabla\psi\propto 1/(r^2+x^2)$ into the bulk far from the pore (as noted by Ref.\cite{yossifon}) and (ii) that the electric field within the pore is divergence-free, such that electro-neutrality is obeyed. From these assumptions it follows that the pore-potential drop $\Delta\psi_{\rm{p}}=-\int_0^L\partial_x\psi \dd x$ is given by
\begin{equation}
    \Delta\psi_{\rm{p}}=\dfrac{\Delta\psi}{1+\dfrac{\pi }{4L}(R_{\rm{b}}+R_{\rm{t}})},
    \label{pot_drop}
\end{equation}
which we derive and verify in SI-4. We note that the edge resistance is negligible in the long-channel limit $L/R_{\rm{b}}\gg 1$, as Eq.(\hyref{pot_drop}) reduces to $\Delta\psi_{\rm{p}}\simeq \Delta\psi$ in this limit. In the geometry of our experiments the reduced potential $\Delta\psi_{\rm{p}}$ given by Eq.(\hyref{pot_drop}) does not only effectively halve the electric current (as $L\approx R_{\rm{b}}+R_{\rm{t}}$ for our experimental geometries), but as noted it will also significantly influence current rectification. As Eq.(\hyref{pot_drop}) accounts for the influence of bulk-edge resistance (red regions in Fig.\hyref{Cartoon_theory}(a)) for both the electric current and (electro-osmotic) flow and we use an effective surface charge $\sigma$ as proxy for the surface-edge resistance, from now on our mathematical analysis will pertain only to the inner-pore region (green-shaded region in Fig.\hyref{Cartoon_theory}(a)) therewith following Ref.\cite{POF}.\\

The electric potential difference over the pore $\Delta\psi_{\rm{p}}$ does not only drive ion fluxes $\bf{j}_\pm(x,r)$ of the cations ($+$) and anions ($-$) but also a fluid flow with a velocity field $\mathbf{u}(x,r)$. The salt flux $\bf{j}_{\rm{s}}=\bf{j}_++\bf{j}_-$ and electro-osmotic flow $\bf{u}$ will be key to understanding the electric current $\bf{j}_{\rm{e}}=\bf{j}_+-\bf{j}_-$. The resulting salt concentration $\rho_{\rm{s}}(x,r)=\rho_+(x,r)+\rho_-(x,r)$ due to the inhomogeneous salt current is of key importance for current rectification, while the space-charge $e\rho_{\rm{e}}(x,r)=e(\rho_+(x,r)-\rho_-(x,r))$ outside the electric double layer is of secondary importance as was shown in Ref.\cite{POF}. The ionic fluxes and concentrations satisfy the Nernst-Planck equations (\hyref{charge})-(\hyref{salt}) that describe diffusion, conduction, and advection,  while the electric potential satisfies the Poisson equation (\hyref{Poisson}) in terms of the electric space charge density $e\rho_{\rm{e}}$. The fluid flow in the low-Reynolds number regime of interest here is given by the Stokes equation (\hyref{Stokes}) that includes an electric body force $-e\rho_{\rm{e}}\nabla\psi$, and the steady-state condition of interest leads to the condition of divergence-free fluxes (\hyref{divergence-free}), which all accumulates into 
\begin{gather}
\mathbf{j}_{\rm{e}}=-D \big(\nabla\rho_{\rm{e}}+\rho_{\rm{s}} \frac{e\nabla\psi}{k_\mathrm{B}T})+\mathbf{u}\rho_{\rm{e}}\label{charge};\\
  \mathbf{j}_{\rm{s}}=-D \big(\nabla\rho_{\rm{s}}+\rho_{\rm{e}} \frac{e\nabla\psi}{k_\mathrm{B}T})+\mathbf{u}\rho_{\rm{s}}\label{salt};\\
\nabla^2\psi=-\frac{e}{\epsilon}\rho_{\rm{e}}; \label{Poisson}\\
\eta\nabla^2 \mathbf{u} - \nabla P-e\rho_{\rm{e}} \nabla\psi=0;\label{Stokes}\\
\nabla\cdot \mathbf{j}_\pm=0, \quad \nabla\cdot \mathbf{u}=0.\label{divergence-free}
\end{gather}
Here Eq.(\hyref{charge}) shows explicitly that the salt concentration $\rho_{\rm{s}}=\rho_++\rho_-$ determines the electric conductivity of the charge current $\bf{j}_{\rm{e}}$. To obtain the pore conductance we consider both reservoirs with a dilute (monovalent) KCl solution of concentration $\rho_{\rm{b}}$, viscosity $\eta=1$ mPa s, a dielectric permittivity $\epsilon=80\epsilon_0$, with $\epsilon_0$ the permittivity of vacuum, and a fixed temperature $T=298$ K, which is constant throughout the system. Deep into the bulk of the base-connected reservoir, $x\ll -L$,  we impose that the K$^+$ and Cl$^-$ concentrations $\rho_\pm=\rho_{\rm{b}}$, $P=P_0$, $\psi=\Delta\psi$, and deep into the tip-connected reservoir, $x\gg 2L$, we impose $\rho_\pm=\rho_{\rm{b}}$, $P=P_0$, and $\psi=0$. Here the reference pressure $P_0=1$ atm. For K$^+$ and Cl$^-$ we use equal diffusion coefficients $D=1$ nm$^2$ ns$^{-1}$, which is somewhat smaller than the bulk diffusion constant at 20 $\degree$C and 0.1 M\cite{difcon1, difcon2}. Such a discrepancy between channel and bulk diffusion constants has been noted in Ref.\cite{pyramidal}.

In thermodynamic equilibrium with vanishing potential drop between the reservoirs $(\Delta\psi=0)$ and vanishing fluxes, the PNPS equations (\hyref{charge})-(\hyref{divergence-free}) reduce to Poisson-Boltzmann theory that describes a diffuse layer of net ionic charge near the surface, known as the electric double layer (EDL) with typical thickness $\lambda_{\rm{D}}=(8\pi\lambda_{\rm{B}}\rho_{\rm{b}})^{-1/2}$, and Bjerrum length $\lambda_{\rm{B}}=e^2/(4\pi\epsilon k_{\rm{B}}T)\simeq 0.7$ nm. Outside this layer all concentrations $\rho_{\pm}(x,r)$ are equal to $\rho_{\rm{b}}$ and there is no electric field, $-\nabla\psi(x,r)\simeq0$. In equilibrium the surface charge density $e\sigma$ obeys the Gouy-Chapman equation, $2\pi\lambda_{\rm{B}}\lambda_{\rm{D}}\sigma=\sinh^{-1}(e\psi_0/2k_{\rm{B}}T)$\cite{physicalchem}. Here $\psi_0$ is the surface potential of the channel wall, which we will use as a fit parameter below, taken to be constant between all geometries and at all concentrations thereby implicitly accounting for a concentration-dependent surface charge $\sigma(\rho_{\rm{b}})$ due to a salt-concentration dependent surface-reaction\cite{surfpotKCl1,surfpotKCl2}. 

For non-vanishing applied potential drops ($\Delta\psi \neq 0$)  not only an electric current $I=2\pi e\bf{\hat{x}}\cdot\int_0^R \bf{j}_{\rm{e}}r\dd r$ (with $\hat{\bf{x}}$ the unit vector along the x-direction) and electro-osmotic flow $Q=2\pi\bf{\hat{x}}\cdot\int_0^R\bf{u}(r)r\dd r$ are driven through the pore, but also a salt current $J=2\pi\bf{\hat{x}}\cdot\int_0^R \bf{j}_{\rm{s}}(r)r\dd r$ where the bulk-excess salt current is primarily a conductive current through the EDL. In steady-state this salt current must be laterally constant to prevent the build up of salt through the pore, $\pi R^2(x)\partial_t \bar{\rho}_{\rm{s}}=-\partial_xJ=0$, where  $\bar{\cdots}=(\pi R^2(x))^{-1} \int_0^{R(x)} \cdots r \dd r$ denotes the cross-sectional average of the salt concentration. The condition of a divergence-free flux (Eq.(\hyref{divergence-free})), which is necessary for a steady-state solution, leads to a differential equation for cross-sectionally averaged salt concentration for $x\in[0,L]$,
\begin{equation}\label{const salt}
     D\partial_x \left(\pi R^2 \partial_x\bar{\rho}_s+ 2\pi R \sigma \frac{e\partial_x\psi}{k_\mathrm{B}T}\right) - Q\partial_x\bar{\rho}_s=0,
\end{equation}
with the electric field $-\partial_x\psi=(\Delta\psi_{\rm{p}}/L) R_{\rm{b}} R_{\rm{t}}/R^2(x)$\cite{POF} and the electro-osmotic flow in a conical channel $Q=-\Delta\psi_{\rm{p}} (\pi R_{\rm{t}}R_{\rm{b}}/L)(\epsilon\psi_0/\eta)$, (as derived beneath Eq.(2) in Ref.\cite{POF}). In Eq.(\hyref{const salt}) the first term represents diffusion of salt through the bulk of the pore, the second term conduction of salt through the EDL, and the third term advection of salt through the bulk of the pore. In a cylinder with constant radius $R$ this differential equation reduces to $D\partial_x^2\bar{\rho}_{\rm{s}}-Q\partial_x\bar{\rho}_{\rm{s}}=0$, which with boundary conditions $\bar{\rho}_{\rm{s}}(0)=\bar{\rho}_{\rm{s}}(L)=2\rho_{\rm{b}}$ has the trivial solution of $\bar{\rho}_{\rm{s}}=2\rho_{\rm{b}}$. Thus for straight pores no current rectification is expected. For a conical geometry, the laterally changing radius $R(x)$ causes lateral variation of conductive currents $D\partial_x (2\pi R(x)\sigma e\partial_x\psi(x) /k_\mathrm{B}T)\neq 0$ which frustrates the formation of a constant $J$ and acts as a non-zero source term. For a negative surface charge, as is typically the case for silica, this source term is negative for $\Delta\psi>0$ and thus the salt concentration in the pore decreases. For $\Delta\psi<0$ this source term is positive, and thus the salt concentration increases. Solving for the cross-sectional average concentration profile $\bar{\rho}_{\rm{s}}(x)$ leads to a non-trivial solution\cite{POF}
    \begin{subequations}
\begin{align}
    \bar{\rho}_{\rm{s}}(x) - 2\rho_{\rm{b}}&=  \frac{\Delta \rho}{\text{Pe}} \Bigg[
     \frac{x}{L} \frac{ R_{\rm{t}}}{R(x)}-\frac{ \exp\left({\displaystyle\frac{x}{L}\frac{ R_{\rm{t}}^2}{  R_{\rm{b}} R(x)}\text{Pe}}\right)-1}{\exp\left({\displaystyle\frac{R_{\rm{t}}}{ R_{\rm{b}}}\text{Pe}}\right)-1}\Bigg] \label{eq:concentrationa}\\
      &= \dfrac{\Delta\rho}{2|\text{Pe}|}\bigg(\frac{ R_{\rm{b}}}{R(x)}\big(1-\frac{x}{L}(1+\frac{R_{\rm{t}}}{ R_{\rm{b}}})\big)\mp 1\bigg) \quad \text{if \ $\pm$Pe} \gg \bigg(\frac{R_{\rm{b}}}{R_{\rm{t}}}\bigg)^2, \label{eq:concentrationb}
    \end{align}
    \label{eq:concentration}
\end{subequations} where the tip P\'eclet number $\text{Pe} \equiv -\Delta\psi_{\rm{p}} (R_{\rm{b}}/ R_{\rm{t}})(\epsilon\psi_0/D\eta)$ and $\Delta \rho \equiv 4(e \Delta \psi_{\rm{p}}/k_{\rm{B}}T)\text{Du} (R_{\rm{b}}/R_{\rm{t}}-1)\rho_{\rm{b}}$ is a measure for the concentration polarization, with tip Dukhin number $\text{Du}=\sigma/(2\rho_{\rm{b}}R_{\rm{t}})$. Note that both Pe and Du carry a sign and the diode polarity stems from the sign of the Dukhin number.

Having obtained the salt-concentration $\bar{\rho}_{\rm{s}}(x)$ in Eq.(\hyref{eq:concentration}) the resulting pore conductance $G(\Delta\psi)=I(\Delta\psi)/\Delta\psi$ is calculated by cross-sectionally integrating Eq.(\hyref{charge}) which results in
\begin{equation}
    G(\Delta\psi)=G_{\rm{b}}(\Delta\psi)\bigg(1+\frac{4\langle\lambda_{\rm{D}}\rangle}{ R_{\rm{b}}+R_{\rm{t}}} \big(\cosh\big(\frac{e\psi_0}{2k_{\rm{B}}T}\big)-1\big)\bigg),
    \label{gen_conductance}
\end{equation}
here $\langle \cdots \rangle = L^{-1}\int_0^L \cdots \dd x$ denotes the lateral average, and the bulk channel conductance is given by $G_{\rm{b}}(\Delta\psi)=4 \langle\bar{\rho}_{\rm{s}}\rangle R_{\rm{t}}R_{\rm{b}}(e^2D/k_{\rm{B}}T) /(4L/\pi+ R_{\rm{b}}+ R_{\rm{t}})$. This bulk conductance also accounts for the in- and outlet resistance by incorporation of Eq.(\hyref{pot_drop}) and depends on $\Delta\psi$ through the potential dependence of $\langle\bar{\rho}_{\rm{s}}\rangle$, which is obtained by numerical integration of Eq.(\hyref{eq:concentration}). We note that Eq.(\hyref{gen_conductance}) obtained from the PNPS equations (\hyref{charge})-(\hyref{divergence-free}) has precisely the form expected from the circuit depicted in Fig.\hyref{Cartoon_theory}(a): it consists of the sum of a bulk and surface (pore) conductance, $G=G_{\rm{b}}+G_{\rm{s}}$ (Fig.\hyref{Cartoon_theory}(a, green)), whereas the tip and base conductances (Fig.(\hyref{Cartoon_theory}(a, red)) stand in series with the pore and lower the total conductance per Eq.(\hyref{pot_drop}). The surface conductance $G_{\rm{s}}=4G_{\rm{b}}\langle\lambda_{\rm{D}}\rangle/( R_{\rm{b}}+R_{\rm{t}}) \big(\cosh(e\psi_0)/(2k_{\rm{B}}T)-1\big)$ will vary with concentration through the dependence of the "channel-weighted" Debye length $\langle\lambda_{\rm{D}}\rangle\simeq(4\pi\lambda_{\rm{B}}\langle\bar{\rho}_{\rm{s}}\rangle)^{-1/2}$. In principle we could include the advective (streaming current) contribution to Eq.(\hyref{gen_conductance}), but its contribution to the surface conductance is proportional to $k_{\rm{B}}T/(4\pi\eta\lambda_{\rm{B}}D (\cosh(e\psi_0/2k_{\rm{B}}T)-1))\ll 10^{-2}$ for all our parameter sets and hence can be neglected\cite{werkhovensoftmatter}.

It is important to note that while the advective contribution to the electric current $I$ can be neglected the advective contribution to the salt current $J$ (Eq.(\hyref{const salt})) is key to current rectification as for ICR the flow rate determines the characteristic voltage $\Delta\psi_{\rm{c}}$, known as the knee voltage for diodes. When $\Delta\psi_{\rm{c}}\gg \Delta\psi$ conductance is Ohmic ($G_0$), while for $\pm\Delta\psi\gg\Delta\psi_{\rm{c}}$ the limiting diode conductance ($G_\pm$) has been reached. From Eq.(\hyref{eq:concentration}) it can be seen that for large flow |Pe|$\gg (R_{\rm{b}}/R_{\rm{t}})^2$ the concentration profile $\bar{\rho}_{\rm{s}}(x) - 2\rho_{\rm{b}}\propto  \Delta\rho/|\text{Pe}|$ is potential independent and hence per Eq.(\hyref{gen_conductance}) this limiting conductance $G_\pm$ has been reached. Hence the characteristic potential $\Delta\psi_{\rm{c}}$ is set by the potential for which Pe=$ (R_{\rm{b}}/R_{\rm{t}})^2$ yielding
\begin{equation}
\Delta\psi_{\rm{c}}=(R_{\rm{b}}/R_{\rm{t}})(D\eta/\epsilon|\psi_0|)[1+(\pi/4L)(R_{\rm{b}}+R_{\rm{t}})],
\label{char_pot}
\end{equation}
where the term in the square brackets of Eq.(\hyref{char_pot}) accounts for edge resistance and can be set to unity in the long-channel limit. While Eq.(\hyref{gen_conductance}) can be used to describe the conductance for arbitrary $\Delta\psi$ by straightforward numerical integration of Eq.(\hyref{eq:concentrationa}), a more convenient closed form for the limiting conductances $G_\pm$ can be found when neglecting the second (surface) term for the electric conductance $G(\Delta\psi)$ Eq.(\hyref{gen_conductance}). This approximation therefore neglects surface conductance entirely and subsequently integrating Eq.(\hyref{eq:concentrationb}) yields
\begin{equation}
\frac{G_{\pm,\rm{b}}}{G_{0,\rm{b}}} = 1+2w\textrm{Du} \bigg[\frac{\log\big(R_{\rm{b}}/R_{\rm{t}}\big)}{R_{\rm{b}}/R_{\rm{t}}-1}-\bigg(\frac{R_{\rm{t}}}{R_{\rm{b}}}\bigg)^{\frac{1\pm1}{2}}\bigg].
    \label{ICR}
\end{equation}
As such, current rectification is defined by the ratio $G_-/G_+=$ ICR where, as before, Du=$\sigma/(2\rho_{\rm{b}}R_{\rm{t}})$ represents the ratio of salt transport in the EDL with respect to salt transport in the bulk, and $w=e D\eta/(k_{\rm{B}}T\epsilon|\psi_0|)$ is the ratio of the ionic to electro-osmotic mobility that quantifies the competition between the rate by which conduction adds ions to the concentration profile $\bar{\rho}_{\rm{s}}(x)$ and electro-osmotic flow removes them. Note that this ratio depends only on electrolyte properties and surface potential and it is not influenced by the geometry whatsoever, being constant ($w\simeq 0.4$) over all our geometries and concentrations. The bracketed term of Eq.(\hyref{ICR}) fully captures the effect of geometry on the concentration profile $\bar{\rho}_{\rm{s}}(x)$. This last term is zero for $R_{\rm{t}} = R_{\rm{b}}$, positive for $G_+$ and negative for $G_-$ and reflects the influence of geometry on diode polarity. Eq.(\hyref{ICR}) also straightforwardly gives a simple and convenient analytic expression for the ICR = $G_-/G_+$.

In the following sections we consider the small and large potential limits of  Eq.(\hyref{gen_conductance}) to interpret our experimental data by first considering the measured (Ohmic) conductance $G_0$ at small potential drops $\Delta\psi\ll k_{\rm{B}}T/e$ and then to describe ICR, which is given by the ratio of conductances $G_-/G_+$ in the limit of large positive (+) or negative (-) potential drops for $\pm\Delta\psi\gg \Delta\psi_{\rm{c}}$.

\subsection{Ohmic conductance}
\begin{figure}[h!]
    \includegraphics[width=0.6\textwidth]{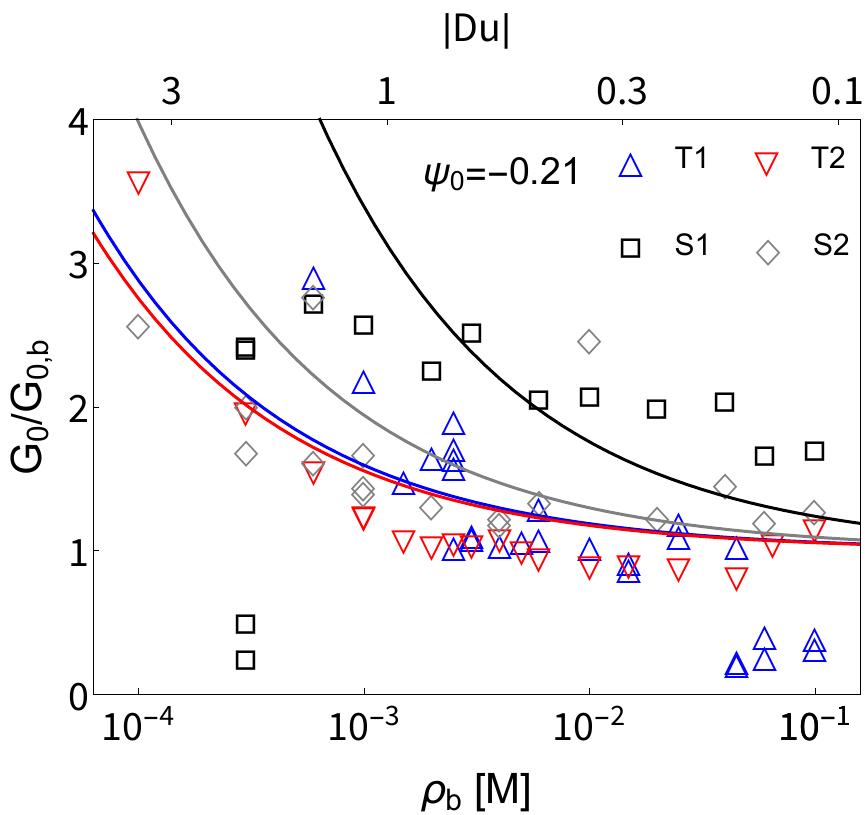}
    \caption{Ohmic conductance $G_0$ in units of the bulk conductance $G_{0,\rm{b}}$ as a function of the bulk concentration $\rho_{\rm{b}}$ (lower axis) and Dukhin number Du$=\sigma/(2\rho_{\rm{b}}R_{\rm{t}})$  (upper axis) with tip radius $R_{\rm{t}}=0.5$ $\mu$m representative for the tapered pores T1 and T2 and straight pores S1, but not for the straight pore S2 with radius $R_{\rm{t}}=1.5 \mu$m. The symbols denote the experimental measurements and the lines with corresponding colors are plotted using Eq.(\hyref{gen_conductance}) with a large surface potential $\psi_0=-0.21$ V (see text). As expected the conductance converges to the bulk conductance at high-concentration while it increases to triple the bulk conductance at low concentrations due to the contribution of surface conductance.}
    \label{fig:Ohm}
\end{figure}

First we consider the Ohmic conductance, $G_0$, which is found at small potential drops  $|\Delta\psi|\ll k_{\rm{B}}T/e$. In this case the laterally averaged concentration equals the bulk concentration $\langle\bar{\rho}_{\rm{s}}\rangle=2\rho_{\rm{b}}$. Hence $G_0$ is given by Eq.(\hyref{gen_conductance}) where the bulk Ohmic conductance $G_{\rm{0,b}}=8 R_{\rm{t}}R_{\rm{b}}(e^2D/k_{\rm{B}}T) /(4L/\pi+ R_{\rm{b}}+ R_{\rm{t}})\rho_{\rm{b}}$ and the surface Ohmic conductance $G_{0,\rm{s}}$ is determined by the equilibrium Debye length $\lambda_{\rm{D}}=(8\pi\lambda_{\rm{B}}\rho_{\rm{b}})^{-1/2}$. In this regime we find that Eq.(\hyref{gen_conductance}) reduces to several well-known results depending on the geometry. The conductance of a long conical channel with negligible surface conductance is retrieved when $L\gg R_{\rm{b}}\textgreater R_{\rm{t}}\gg \lambda_{\rm{D}}$, \cite{POF} the Hall conductance of a thin cylindrical pore with negligible surface conductance is retrieved when $L\simeq R_{\rm{b}}= R_{\rm{t}}\gg\lambda_{\rm{D}}$ \cite{hall1975access}, and the conductance of a long cylindrical channel \cite{werkhovensoftmatter} is obtained when $L\gg R_{\rm{b}}=R_{\rm{t}}\textgreater \lambda_{\rm{D}}$. Hence Eq.(\hyref{gen_conductance}) extends these classical results to short pores with unequal tip and base radii. Figure \hyref{fig:Ohm} shows the experimental Ohmic conductance obtained as $G_0=(I(0.05$ $\rm{V})-I(-0.05$ $\rm{V}))/0.1$ $\rm{V}$ together with our theoretical model Eq.(\hyref{gen_conductance}) for all four channels T1, T2, S1 and S2, where we use $\psi_0=-0.21$ V as it provides the closest match to the data for all concentrations and geometries. Note that the classical long-channel theory that neglects entrance resistance through Eq.(\hyref{pot_drop}) would overestimate the conductance by a factor of about two for our parameters, as the effective potential drop is nearly halved ($0.46 \textless \Delta\psi_{\rm{p}}/\Delta\psi \textless 0.68 $) in our experimental geometries. This reduction of the effective potential drop highlights the importance of edge resistances.
 It should be emphasized that the experimental conductance in Fig. \hyref{fig:Ohm} is normalized by the theoretical bulk conductance $G_{0,\rm{b}}$ from Eq.(\hyref{gen_conductance}), which is determined by both pore geometry and electrolyte properties. At high concentrations this representation highlights that the surface conductance is negligible, as the conductance in units of $G_{0,\rm{b}}$ approaches unity and the experimental data for different geometries collapse into a single curve. Note that Eq.(\hyref{gen_conductance}) properly accounts for the in- and outlet resistance at the highest concentration. Some of the deviation between experimental data and theory are attributed to morphological changes due to clogging over the course of the experiments (see for instance SEM image of T1 after the experiments in SI-5). \\

At low concentrations, $\rho_{\rm{b}}\textless 1$ mM, both the theoretical and experimental conductance rapidly increase as the concentration decreases, which is due to the surface conductance $G_{0,\rm{s}}$ contribution increasing with the increasing Debye screening length. We observe that the concentration at which surface conductance becomes comparable to bulk conductance occurs when the tip Dukhin number approaches unity, $G_{0,\rm{s}}/G_{0,\rm{b}}\propto$ Du= $\sigma/(2\rho_{\rm{b}}R_{\rm{t}})\simeq 1$, which for both T1 and T2 occurs near $\rho_{\rm{b}}\simeq 2$ mM for $\psi_0=-0.21$ V. The experimental variation of conductance with concentration roughly scales as the inverse square of the concentration, increasing by a factor of three when the concentration is decreased by a factor ten. This scaling can only be understood by using a concentration-independent $\psi_0$ as opposed to a concentration-independent surface charge density $\sigma$. With constant $\sigma$ the Dukhin number scales as Du $\propto \rho_{\rm{b}}^{-1}$ and introduces a surface conductance which varies by orders of magnitude in our concentration range, which is not observed. Instead, when using a constant $\psi_0$ the surface charge scales as $\sigma\propto 1/\lambda_{\rm{D}}$ according to the Gouy-Chapman equation, in which case the proper scaling Du $\propto \rho_{\rm{b}}^{-1/2}$ is immediately obtained. The existence of a constant surface potential implies that a chemical reaction is responsible for the surface charge varying with salt concentration. However, while Fig. \hyref{fig:Ohm} indeed shows that the experimental conductance qualitatively follows the inverse square-root scaling, there is a minor quantitative deviation. We attribute this chiefly to charge-regulation beyond the constant-potential model of the silica-water interface\cite{regulation}, which could introduce variations in $\psi_0$ by a factor $\sim 3$ in the range [$10^{-1}-10^{-4}$] M for silica \cite{surfpotKCl1,surfpotKCl2, silica1}. We have not included this concentration effect as there is no unanimous quantitative measurement of charge-regulation for silica \cite{surfpotKCl1,surfpotKCl2, kosmulski, silica1, regulation} and as to prevent overfitting.

\subsection{Ion current rectification}
\begin{figure}[h!]
    \includegraphics[width=1\textwidth]{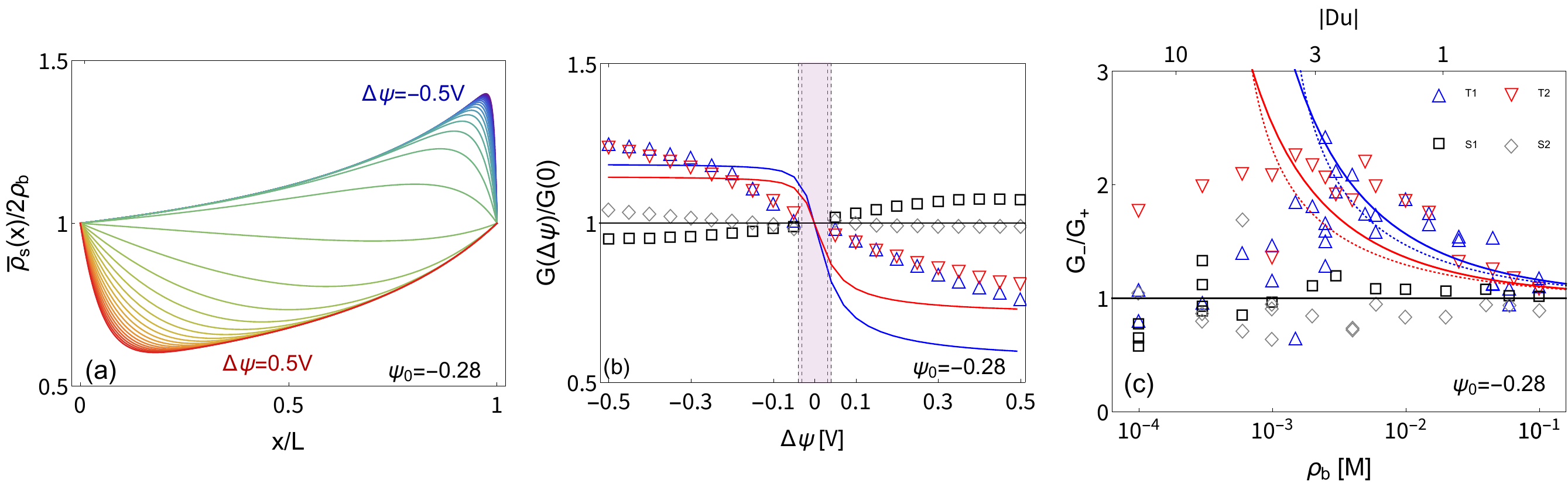}
  \caption{(a) Concentration profiles $\bar{\rho}_{\rm{s}}(x)/2\rho_{\rm{b}}$ for geometry T1 (see text) at a bulk concentration $\rho_{\rm{b}}=6$ mM as calculated by Eq.(\hyref{eq:concentration}) for $\Delta\psi$ between $-0.5$ and $0.5$ V with step-size of $0.03$ V. Depletion occurs for positive potentials (red) while concentration increases at negative potentials (blue). (b) Conductance normalized by the conductance at $\Delta\psi=0$ at varying potentials and the same concentration as in (a) where the different symbols represent the tapered (T1,T2) and straight (S1,S2) channels, and lines are plotted using the combination of Eq.(\hyref{gen_conductance}) and Eq.(\hyref{eq:concentration}a) both with a surface potential of $\psi_0$ = $-0.28$ V. Vertical lines demarcating shaded area are placed at the characteristic voltage $\Delta\psi_{\rm{c}}\simeq \pm 0.08$ V with color corresponding to the respective geometry. Most variation of the experimental conductance occurs within the shaded region, after which the conductance converges to the limiting conductances $G_\pm$. (c) Current rectification given by the ratio $G_{-}/G_{+}$ which for experiments is taken as $G(\pm 0.5$ V) with varying concentration (lower axis) and Dukhin number for $R_{\rm{t}}=0.5$ $\mu$m (upper axis). Solid lines are plotted with our approximation Eq.(\hyref{ICR}) that neglects surface conductance and dotted lines are from the full solution using the combination of Eq.(\hyref{gen_conductance}) and Eq.(\hyref{eq:concentration}a) both using $\psi_0=-0.28$ V. Peak experimental current rectification is reached near Du $\simeq$ 3 while solid lines grow monotonically with Du.}
  \label{fig:ICR}
  \end{figure}

We now turn to large potential drops where $\Delta\psi\gg\Delta\psi_{\rm{c}}$ (Eq.(\hyref{char_pot})) where we observe significant current rectification for tapered pores T1 and T2 and the conductance has converged to its limiting value $G_\pm$ (Eq.(\hyref{ICR}). Current rectification in conical pores is well established to be due to the salt concentration in the pore changing with the applied potential\cite{rect1,rect2,rect3, POF} as described in the theoretical section. The dependence of the laterally averaged salt concentration $\langle\bar{\rho_{\rm{s}}}\rangle$ on $\Delta\psi$ as in Eq.(\hyref{eq:concentration}) in conjunction with our expression for the conductance Eq.(\hyref{gen_conductance}) immediately result in a voltage-dependent conductance. In Fig.\hyref{fig:ICR}(a) we use Eq.(\hyref{eq:concentration}) to plot the salt concentration profiles $\rho_{\rm{s}}(x)/(2\rho_{\rm{b}})$ in the conical pore T1 for $\Delta\psi$ between $-0.5$ V and $0.5$ V and a concentration $\rho_{\rm{b}}=6$ mM. It can be seen that for negative voltages there is a build-up of ions, while for positive voltage drops the pore becomes depleted. This increase and decrease, characteristic of the conical geometry, is highly dependent on the applied voltage, but converges to a limiting concentration profile for which inhomogeneous conduction is balanced by advection. These limiting concentration profiles can deviate up to $50\%$ from the bulk concentration and in turn significantly modulate the voltage-dependent conductance as can be seen in Fig.\hyref{fig:ICR}(b). Here we compare $G(\Delta\psi)$ from Eq.(\hyref{gen_conductance}) with the experimental conductance normalized by the experimental Ohmic conductance $G(\Delta\psi=0)$. We observe two plateaus of high and low conductance for the theoretical curves at large negative and positive voltages for the tapered pores. The transition between the two regimes occurs at the borders of the shaded region $|\Delta\psi|\leq\Delta\psi_{\rm{c}}\simeq0.05$ V beyond which the conductance quickly converges to the limiting conductance $G_\pm$. In SI-7 we show more plots at different concentrations for comparison, which all show the same typical S-shaped curve with the exception for curves at $\rho_{\rm{b}}\textless 1$ mM for which the experimental variation is larger due to leakage currents as discussed in the experimental section.\\
In Fig. \hyref{fig:ICR}(c) we plot the experimental ICR against concentration and tip Dukhin number, together with results based on both Eq.(\hyref{ICR}) (solid) and the combination of Eq.(\hyref{gen_conductance}) with Eq.(\hyref{eq:concentrationa}) (dashed) using a fitted surface potential $\psi_0=-0.28$ V in both cases. Theoretical and experimental ICR obey the same inverse square root scaling $G_-/G_+\propto\rho_{\rm{b}}^{-1/2}$ as was also observed for Ohmic conductance, which is again due to the concentration dependence of Du at fixed surface potential $\psi_0$. Interestingly, we find that Eq.(\hyref{ICR}) is a good approximation for the combination of Eq.(\hyref{gen_conductance}) with Eq.(\hyref{eq:concentrationa}).  The unexpected quality of our approximation Eq.(\hyref{ICR}) is a result of a cancellation of errors: an increase of Ohmic conduction due to surface conductance decreases ICR while the variation of surface conductance with concentration increases ICR.\\ 
At low concentrations the inverse square root dependence of ICR breaks down around $\rho_{\rm{b}}\simeq2$ mM, where the experimental ICR peaks while our theory predicts that ICR should keep increasing with decreasing concentration. Such a peak in ICR has been previously observed in long micro-channels\cite{cervera_exp_micropores_and_saturation,siwy_saturation,rect1} and assigned to the emergence of EDL overlapping at low-concentrations. Here the concentration depletion in the pore only allows for minor EDL overlap at the tip, as the screening length is at least an order of magnitude smaller than the tip size in our experiments. Other theoretical works predict a peak in ICR at $1$<Du<$10$ \cite{Poggioli2019, Cengio2019}, which is attributed to salt transport in the EDL dominating the total salt transport, an effect that is not captured in our model. In SI-9 we plot the pore selectivity as defined in Ref.\cite{Cengio2019} for our tapered geometries with $\psi_0=-0.21$ V and find a maximum near $\rho_{\rm{b}}=2$ mM, in line with our experiments.

\subsection{Discussion of the large surface potential}
Both conductance $G_0$ and current rectification $G_-/G_+$ are visually fitted using $\psi_0$ as the only fit parameter which we keep constant for all geometries and concentrations, yielding $\psi_0=-0.21$ V for $G_0$ and $\psi_0=-0.28 $ V for $G_-/G_+$. However, the surface chemistry of the silica interface is well studied, and a much lower surface potential between -0.03 V and -0.1 V is expected in our experimental conditions\cite{kosmulski}. While surface potentials may vary quite significantly between different measurement methods, protocols and even subsequent measurements\cite{kosmulski}, a discrepancy that exceeds 0.1 V is excessive. From the Gouy-Chapman equation we find that a pore with a surface potential between -0.21 and -0.28 V would contain approximately 15-60 times more charge than for a typical literature surface potential of $-0.07$ V. Such a large discrepancy cannot be explained by subtle experimental factors, and it has to be assumed that this deviation stems from the theoretical model not including all of the key physics. \\
In our analysis we exclude the charge on the planar membrane outside the pore, effectively neglecting an edge EDL conductance $G^{\rm{base}}_{\rm{s}}$ and $G^{\rm{tip}}_{\rm{s}}$ as depicted in Fig. \ref{Cartoon_theory}(a). Other authors have noted that this region on the outside of the pore can contribute to both the Ohmic conductance \cite{Lee2012, pyramidal} and the current rectification \cite{Ma2018, yossifon} for thin pores with $R_{\rm{b}}/L\leq 1$. The charge on the outside of the membrane not only increases the edge conductance as noted by\cite{Lee2012, Bocquet2020} but surprisingly can also induce excess ICR\cite{yossifon}. This excess ICR is due to a radial electric field driving an inhomogeneous salt current through the EDL ouside of the pore, leading to accumulation/depletion of salt in the reservoir as we demonstrate in SI-6. While excess ICR and excess conductance both occur in the EDL outside of the pore, they are distinct phenomena whose scaling and characteristic length scales may differ qualitatively, explaining why our experimental $G_0$ and $G_+/G_-$ are better accounted for with two different surface potentials, $\psi_0=-0.21$ V and $\psi_0=-0.28$ V, respectively.\\

\begin{figure}[h!]
	\includegraphics[width=.4\textwidth]{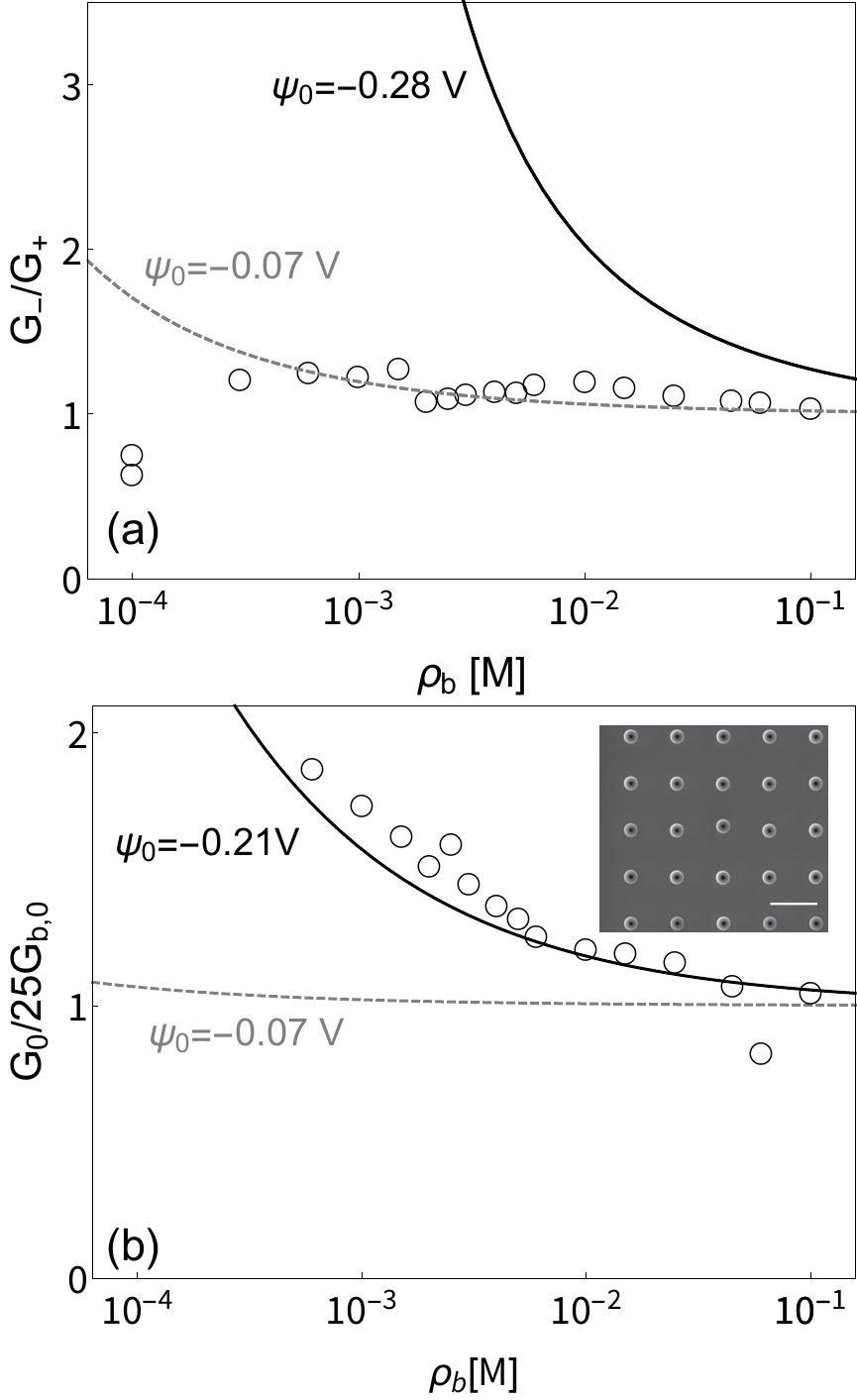}
	\caption{ (a) Current rectification $G_{-}/G_{+}$ as function of concentration (lower axis) for an array of 25 pores. The symbols denote the experimental measurements and the lines are plotted using Eq.(\hyref{ICR}) using a surface potential of $\psi_0=-0.28$ V (black, continuous) and $\psi_0=-0.07$ V (gray, dashed). No current rectification is found in the experiment, in line with a pore with a surface potential of $\psi_0=-0.07$ V. (b) Dimensionless Ohmic conductance $G_0$ in units of the bulk conductance $G_{0,\rm{b}}$ in the same representation as in Fig.\hyref{fig:Ohm} with lines plotted using Eq.(\hyref{gen_conductance}). The measured conductance corresponds essentially to 25 times the conductance of a single pore with $\psi_0=-0.21$ V. The inset of (b) shows an SEM image of the array directly after fabrication, the scale bar is 10 $\mu$m.}
	\label{array}
\end{figure}
Considering that the charge on the outside of the membrane can contribute to both ICR and increased conductance we suggest to assign the excess charge from the large, fitted, surface potential $\psi_0$ to the charge located on the planar membrane outside of the pore. We estimate the charged-surface area $A_{\rm{out}}$ outside the pore contributing to current-rectification as $A_{\rm{out}} = A_{\rm{pore}}\big((\sigma_{\rm{app}}/\sigma_{\rm{lit}})-1\big)$, where $A_{\rm{out}}$ is the (circular) area outside of the pore contributing to entrance-surface conductance, $A_{\rm{pore}}$ is the surface area of the conical pore, $\sigma_{\rm{app}}=\sinh(\psi_0 e/2k_{\rm{B}}T)/(2\pi \lambda_{\rm{B}}\lambda_{\rm{D}})$ is the apparent (Gouy-Chapman) surface charge density resulting from the fitted surface potential (with $\psi_0=-0.21$ V for Ohmic conductance and $\psi_0=-0.28$ V for ICR) and $\sigma_{\rm{lit}}=\sinh(-0.07$ $\mathrm{V} (e/2k_{\rm{B}}T))/(2\pi \lambda_{\rm{B}}\lambda_{\rm{D}})$ is the surface charge density as calculated from a literature surface potential of 0.07 V. With these values we find that the outer-membrane EDL within a radius of about 15.0 $\mu$m from the pore center contributes to ICR, while a shorter radius of only 7.4 $\mu$m contributes to Ohmic conductance. This latter value closely corresponds to the Dukhin length $\sigma/(2\rho_{\rm{b}})\simeq$ 7 $\mu$m at 10$^{-4}$ M which was predicted by Refs.\cite{Lee2012, Bocquet2020} to set the length-scale for (Ohmic) outer-membrane surface conductance. \\

To verify this larger surface contribution of 15 $\mu$m for ICR we derive a solution for the concentration polarization far from the pore in SI-6, and find that the concentration profile in the bulk obeys a long-ranged inverse square law decay $\rho_{\rm{s}}(x, r)\propto (r^2+x^2)^{-1}$ like the electric field with a prefactor proportional to the inverse aspect ratio $R_{\rm{b}}/L$. This prefactor indicates that outer-membrane concentration polarization only occurs for short pores, while the inverse square decay indicates that the concentration profile decays over lengths much larger than the radius. Both these observations support the hypothesis that surface charge far from the pore can contribute to current rectification for low-aspect ratio pores. Unfortunately we were unable to construct a theory simultaneously accounting for concentration polarization in- and outside the pore, while numerical (COMSOL) calculations of the full PNPS Eqs.(\hyref{charge})-(\hyref{divergence-free}) proved unstable. To experimentally test our hypothesis we construct a 5$\times$5 array of conical pores with dimensions $R_{\rm{t}} \simeq 0.35$ $\mu$m, $R_{\rm{b}} \simeq $1.4 $\mu$m and $L\simeq 2$ $\mu$m with a spacing of $10$ $\mu$m between the pore centers ($\approx$ 10$^6$ pores/cm$^2$). In Fig.\hyref{array}(a) we show the ICR calculated with Eq.(\hyref{ICR}) using the literature surface potential $\psi_0=-0.07$ V (dashed line) and the large surface potential obtained from the single pore fitted with $\psi_0=-0.28$ V (solid line) together with the experimental data for the array (symbols). We observe that ICR is greatly reduced for the array, virtually disappearing over the whole concentration range. In contrast the Ohmic conductance of the array is essentially 25 times the single pore conductance given by Eq.(\hyref{gen_conductance}) as shown in Fig.\hyref{array}(b). Considering that the single pore results are based on the fitted value of $\psi_0$ = -0.21 V, the surface conductance therefore remains excessively large compared to the expectation at a literature surface potential of $\psi_0=-0.07$ V (solid line). In line with our hypothesis, these observations therefore surprisingly imply that the charge on the outside of the membrane contributes over a smaller range to conductance than to ICR, so that interference only occurs for the latter at this spacing. 

\subsection{Conclusion}
To summarize, we have presented experimental and theoretical results on ion current rectification (ICR) in tapered micropores connecting aqueous KCl solutions, leading to three main results. \\ 
(i) We demonstrate the existence of ICR in conical micropores fabricated in crystalline silicon membranes without further chemical modification at KCl concentrations where the (bulk) electrolyte screening length is much smaller than the pore size, and which is absent in straight cylindrical pores. (ii) We derive an expression for the conductance of short conical pores accounting for both the EDL within the channel as well as the edge resistance at the tip and base of the pore. These edge resistances approximately halve the Ohmic conductance in our experimental geometries. Our expression (Eq.(\hyref{gen_conductance})) reverts to the Hall conductance in case of thin cylindric pores\cite{hall1975access}, conical conductance in case of long cones\cite{POF}, and the well-known conductance of straight cylinders with EDL's for large aspect ratio channels\cite{werkhovensoftmatter}. We find an expression for the characteristic voltage at which current rectification occurs, $\Delta\psi_{\rm{c}}\simeq 0.05$ V in our geometries, and find a new closed-form expression (Eq. (\ref{ICR})) for the limiting ICR at large potential drops. While, like other authors, we find that rectification scales with the Dukhin number, our expression contains two new dimensionless terms: the ratio $w$ of the ionic and electro-osmotic (fluid) mobility and a term describing the influence on geometry which solely depends on the tip-to-base ratio. Using two different surface potentials ($\psi_{0}$ = -0.21 V for Ohmic conductance and $\psi_0=-0.28$ V) our theory closely matches our experimental results for all but the lowest concentrations and largest potential drops. In this regime of extreme depletion minor EDL overlap occurs at the tip, invalidating the starting assumption of non-overlapping EDL's in our theory.
\\
(iii) Finally we discuss the physical interpretation of the surface potential $\psi_{0}$ which is our sole fit parameter. Our fitted surface potential is excessively large compared to literature values and should not be interpreted as the actual potential but rather as an apparent surface potential. This apparent potential is inflated by the contribution of charge on the outside of the membrane, a region explicitly excluded from our theoretical description. We estimate from the fitted $\psi_0$ that charge on the membrane surface within about $7.4$ $\mu$m of the pore contributes to (Ohmic) conductance at low potentials and within $15.0$ $\mu$m to ICR at larger potentials. We test this hypothesis by fabricating an array of 25 pores with a $10$ $\mu$m separation of the pore centers ($\approx$ 10$^6$ pores/cm$^2$) with no overlap of the low potential (Ohmic) interaction length and large overlap of the high potential (ICR) interaction length. While we observe no pore-pore interference for Ohmic conduction at low potentials, we indeed find that ICR vanishes in the array, in agreement with our hypothesis. The interaction length for Ohmic conduction is known to be set by the Dukhin length\cite{Lee2012,Bocquet2020} while for ICR we show that a long-ranged, inverse-square-distance decay determines the pore-pore interaction, in line with our experimental observations. For thin membranes this apparent contribution of the charge on the outside of the membrane to both surface conductance and ICR may be beneficial for single pores, however these contributions could be detrimental in densely packed arrays that would be desirable for applications. Further investigation of the interaction length for outer-membrane conductance and ICR with different pore densities is therefore particularly relevant. Here, the presented crystalline silicon membranes provide an attractive platform compatible with conventional fabrication methods for e.g. creating homogeneous pore walls through wet etching or engineering of the pore behavior by modification of the outer-membrane surface.

\subsection{Methods}

\subsubsection{Pore fabrication}

Crystalline silicon membranes were purchased from Norcada (SM5200N, thickness 2 $\mu$m $\pm$0.5 $\mu$m). Pores were milled in the membrane with a focused ion beam (FEI Helios Nanolab 600, Ga ions), with tapered pores created by milling concentric circles. Scanning electron miscroscopy images were made in the same system.

\subsubsection{Current measurements}

I-V measurements were conducted in a homemade cell consisting of 2 reservoirs, with the membrane fixed in between (as shown in SI-2). The reservoirs are filled with aqueous solutions with equal concentration of KCl (99.99\% from Sigma-Aldrich, in MilliQ\textregistered\ ultrapure water). Measurements are done using a Ag/AgCl wire electrode in each reservoir, connected to a potentiostat (CH760E) in a 2-electrode configuration with the working electrode facing the large opening (base) for the tapered channels. Quasi-static I-V curves were recorded using staircase voltammetry from -1 V to 1 V and back, with potential steps of 0.05 V and with the system set to rest for 150 s at -1 V before starting the measurement. Each potential is maintained for a period of 10 s, with the current recorded for the last 0.5 s. Data shown in Figure \ref{Fig2} is obtained from averaging 1-4 cycles.

\subsubsection{Electrode preparation}

Ag/AgCl wires were fabricated following the protocol in reference \cite{Inamdar2009} by immersing Ag wires (0.35 mm diameter, 99.99\% ) in 0.1 M nitric acid to remove the native oxide and then rinsed with ultrapure water. The cleaned wire was using as a working electrode in a 3-electrode setup with a platinum counter,- and quasi reference electrode in an aqueous solution of 3 M KCl. The Ag wire was coated with AgCl by applying 2 V vs. the Pt wire QRE for 10 minutes.

\subsection{Conflicts of interest}

There are no conflicts to declare.

\subsection{Acknowledgements}

 This work is part of the research program at the Netherlands Organisation for Scientific Research (NWO). We would like to acknowledge R. Brakkee and D. Ursem for contributions to the project, and D. Koletzki for fabrication of the setup. This work is part of the D-ITP consortium, a program of the Netherlands Organisation for Scientific Research (NWO) that is funded by  the  Dutch  Ministry  of  Education,  Culture  and  Science (OCW).


\begin{mcitethebibliography}{61}
\providecommand*\natexlab[1]{#1}
\providecommand*\mciteSetBstSublistMode[1]{}
\providecommand*\mciteSetBstMaxWidthForm[2]{}
\providecommand*\mciteBstWouldAddEndPuncttrue
  {\def\EndOfBibitem{\unskip.}}
\providecommand*\mciteBstWouldAddEndPunctfalse
  {\let\EndOfBibitem\relax}
\providecommand*\mciteSetBstMidEndSepPunct[3]{}
\providecommand*\mciteSetBstSublistLabelBeginEnd[3]{}
\providecommand*\EndOfBibitem{}
\mciteSetBstSublistMode{f}
\mciteSetBstMaxWidthForm{subitem}{(\alph{mcitesubitemcount})}
\mciteSetBstSublistLabelBeginEnd
  {\mcitemaxwidthsubitemform\space}
  {\relax}
  {\relax}

\bibitem[Fau(2019)]{Faucher2019}
{Critical Knowledge Gaps in Mass Transport through Single-Digit Nanopores: A
  Review and Perspective}. \emph{The Journal of Physical Chemistry C}
  \textbf{2019}, \emph{123}, 21309--21326\relax
\mciteBstWouldAddEndPuncttrue
\mciteSetBstMidEndSepPunct{\mcitedefaultmidpunct}
{\mcitedefaultendpunct}{\mcitedefaultseppunct}\relax
\EndOfBibitem
\bibitem[Bocquet(2020)]{Bocquet2020}
Bocquet,~L. {Nanofluidics coming of age}. \emph{Nature Materials}
  \textbf{2020}, \emph{19}, 254--256\relax
\mciteBstWouldAddEndPuncttrue
\mciteSetBstMidEndSepPunct{\mcitedefaultmidpunct}
{\mcitedefaultendpunct}{\mcitedefaultseppunct}\relax
\EndOfBibitem
\bibitem[Nishizawa \latin{et~al.}(1995)Nishizawa, Menon, and
  Martin]{Nishizawa1995}
Nishizawa,~M.; Menon,~V.~P.; Martin,~C.~R. {Metal Nanotubule Membranes with
  Electrochemically Switchable Ion-Transport Selectivity}. \emph{Science}
  \textbf{1995}, \emph{268}, 700--702\relax
\mciteBstWouldAddEndPuncttrue
\mciteSetBstMidEndSepPunct{\mcitedefaultmidpunct}
{\mcitedefaultendpunct}{\mcitedefaultseppunct}\relax
\EndOfBibitem
\bibitem[Vlassiouk \latin{et~al.}(2008)Vlassiouk, Smirnov, and
  Siwy]{Vlassiouk2008}
Vlassiouk,~I.; Smirnov,~S.; Siwy,~Z. {Ionic Selectivity of Single
  Nanochannels}. \emph{Nano Letters} \textbf{2008}, \emph{8}, 1978--1985\relax
\mciteBstWouldAddEndPuncttrue
\mciteSetBstMidEndSepPunct{\mcitedefaultmidpunct}
{\mcitedefaultendpunct}{\mcitedefaultseppunct}\relax
\EndOfBibitem
\bibitem[Daiguji \latin{et~al.}(2005)Daiguji, Oka, and Shirono]{Daiguji2005}
Daiguji,~H.; Oka,~Y.; Shirono,~K. {Nanofluidic Diode and Bipolar Transistor}.
  \emph{Nano Letters} \textbf{2005}, \emph{5}, 2274--2280\relax
\mciteBstWouldAddEndPuncttrue
\mciteSetBstMidEndSepPunct{\mcitedefaultmidpunct}
{\mcitedefaultendpunct}{\mcitedefaultseppunct}\relax
\EndOfBibitem
\bibitem[Kalman \latin{et~al.}(2008)Kalman, Vlassiouk, and Siwy]{Kalman2008}
Kalman,~E.~B.; Vlassiouk,~I.; Siwy,~Z.~S. {Nanofluidic Bipolar Transistors}.
  \emph{Advanced Materials} \textbf{2008}, \emph{20}, 293--297\relax
\mciteBstWouldAddEndPuncttrue
\mciteSetBstMidEndSepPunct{\mcitedefaultmidpunct}
{\mcitedefaultendpunct}{\mcitedefaultseppunct}\relax
\EndOfBibitem
\bibitem[Rees \latin{et~al.}(2015)Rees, Anderson, Privman, Bau, and
  Venton]{biochem1}
Rees,~H.~R.; Anderson,~S.~E.; Privman,~E.; Bau,~H.~H.; Venton,~B.~J. Carbon
  nanopipette electrodes for dopamine detection in Drosophila. \emph{Analytical
  chemistry} \textbf{2015}, \emph{87}, 3849--3855\relax
\mciteBstWouldAddEndPuncttrue
\mciteSetBstMidEndSepPunct{\mcitedefaultmidpunct}
{\mcitedefaultendpunct}{\mcitedefaultseppunct}\relax
\EndOfBibitem
\bibitem[Wang \latin{et~al.}(2017)Wang, Wang, and Mirkin]{biochem2}
Wang,~Y.; Wang,~D.; Mirkin,~M.~V. Resistive-pulse and rectification sensing
  with glass and carbon nanopipettes. \emph{Proceedings of the Royal Society A:
  Mathematical, Physical and Engineering Sciences} \textbf{2017}, \emph{473},
  20160931\relax
\mciteBstWouldAddEndPuncttrue
\mciteSetBstMidEndSepPunct{\mcitedefaultmidpunct}
{\mcitedefaultendpunct}{\mcitedefaultseppunct}\relax
\EndOfBibitem
\bibitem[Vlassiouk \latin{et~al.}(2009)Vlassiouk, Kozel, and Siwy]{biochem3}
Vlassiouk,~I.; Kozel,~T.~R.; Siwy,~Z.~S. Biosensing with nanofluidic diodes.
  \emph{Journal of the American Chemical Society} \textbf{2009}, \emph{131},
  8211--8220\relax
\mciteBstWouldAddEndPuncttrue
\mciteSetBstMidEndSepPunct{\mcitedefaultmidpunct}
{\mcitedefaultendpunct}{\mcitedefaultseppunct}\relax
\EndOfBibitem
\bibitem[Piruska \latin{et~al.}(2010)Piruska, Gong, Sweedler, and
  Bohn]{biochem4}
Piruska,~A.; Gong,~M.; Sweedler,~J.~V.; Bohn,~P.~W. Nanofluidics in chemical
  analysis. \emph{Chemical Society Reviews} \textbf{2010}, \emph{39},
  1060--1072\relax
\mciteBstWouldAddEndPuncttrue
\mciteSetBstMidEndSepPunct{\mcitedefaultmidpunct}
{\mcitedefaultendpunct}{\mcitedefaultseppunct}\relax
\EndOfBibitem
\bibitem[Liu \latin{et~al.}(2022)Liu, Zhang, Cheng, Yang, Li, Xu, Qu, Liang,
  Cheng, and Li]{biochem5}
Liu,~Z.; Zhang,~S.; Cheng,~M.; Yang,~L.; Li,~G.; Xu,~W.; Qu,~H.; Liang,~F.;
  Cheng,~J.; Li,~H. Highly enantioselective recognition of S-ibuprofen by a
  host--guest induced chiral nanochannel. \emph{Analyst} \textbf{2022},
  \emph{147}, 1803--1807\relax
\mciteBstWouldAddEndPuncttrue
\mciteSetBstMidEndSepPunct{\mcitedefaultmidpunct}
{\mcitedefaultendpunct}{\mcitedefaultseppunct}\relax
\EndOfBibitem
\bibitem[van~der Heyden \latin{et~al.}(2006)van~der Heyden, Bonthuis, Stein,
  Meyer, and Dekker]{VanderHeyden2006}
van~der Heyden,~F. H.~J.; Bonthuis,~D.~J.; Stein,~D.; Meyer,~C.; Dekker,~C.
  {Electrokinetic Energy Conversion Efficiency in Nanofluidic Channels}.
  \emph{Nano Letters} \textbf{2006}, \emph{6}, 2232--2237\relax
\mciteBstWouldAddEndPuncttrue
\mciteSetBstMidEndSepPunct{\mcitedefaultmidpunct}
{\mcitedefaultendpunct}{\mcitedefaultseppunct}\relax
\EndOfBibitem
\bibitem[Xiao \latin{et~al.}(2019)Xiao, Jiang, and Antonietti]{Xiao2019}
Xiao,~K.; Jiang,~L.; Antonietti,~M. {Ion Transport in Nanofluidic Devices for
  Energy Harvesting}. \emph{Joule} \textbf{2019}, \emph{3}, 2364--2380\relax
\mciteBstWouldAddEndPuncttrue
\mciteSetBstMidEndSepPunct{\mcitedefaultmidpunct}
{\mcitedefaultendpunct}{\mcitedefaultseppunct}\relax
\EndOfBibitem
\bibitem[Wang \latin{et~al.}(2012)Wang, Kvetny, Liu, Brown, Li, and
  Wang]{wang_hysteresis}
Wang,~D.; Kvetny,~M.; Liu,~J.; Brown,~W.; Li,~Y.; Wang,~G. Transmembrane
  potential across single conical nanopores and resulting memristive and
  memcapacitive ion transport. \emph{Journal of the American Chemical Society}
  \textbf{2012}, \emph{134}, 3651--3654\relax
\mciteBstWouldAddEndPuncttrue
\mciteSetBstMidEndSepPunct{\mcitedefaultmidpunct}
{\mcitedefaultendpunct}{\mcitedefaultseppunct}\relax
\EndOfBibitem
\bibitem[Wang and Wang(2016)Wang, and Wang]{wang_hysteresis2}
Wang,~D.; Wang,~G. Dynamics of ion transport and electric double layer in
  single conical nanopores. \emph{Journal of Electroanalytical Chemistry}
  \textbf{2016}, \emph{779}, 39--46\relax
\mciteBstWouldAddEndPuncttrue
\mciteSetBstMidEndSepPunct{\mcitedefaultmidpunct}
{\mcitedefaultendpunct}{\mcitedefaultseppunct}\relax
\EndOfBibitem
\bibitem[Han \latin{et~al.}(2022)Han, Oh, and Chung]{iontronics_review}
Han,~S.~H.; Oh,~M.-A.; Chung,~T.~D. Iontronics: Aqueous ion-based engineering
  for bioinspired functionalities and applications. \emph{Chemical Physics
  Reviews} \textbf{2022}, \emph{3}, 031302\relax
\mciteBstWouldAddEndPuncttrue
\mciteSetBstMidEndSepPunct{\mcitedefaultmidpunct}
{\mcitedefaultendpunct}{\mcitedefaultseppunct}\relax
\EndOfBibitem
\bibitem[Siria \latin{et~al.}(2017)Siria, Bocquet, and Bocquet]{Siria2017}
Siria,~A.; Bocquet,~M.-L.; Bocquet,~L. {New avenues for the large-scale
  harvesting of blue energy}. \emph{Nature Reviews Chemistry} \textbf{2017},
  \emph{1}, 0091\relax
\mciteBstWouldAddEndPuncttrue
\mciteSetBstMidEndSepPunct{\mcitedefaultmidpunct}
{\mcitedefaultendpunct}{\mcitedefaultseppunct}\relax
\EndOfBibitem
\bibitem[Hou \latin{et~al.}(2011)Hou, Guo, and Jiang]{hou_diode}
Hou,~X.; Guo,~W.; Jiang,~L. Biomimetic smart nanopores and nanochannels.
  \emph{Chemical Society Reviews} \textbf{2011}, \emph{40}, 2385--2401\relax
\mciteBstWouldAddEndPuncttrue
\mciteSetBstMidEndSepPunct{\mcitedefaultmidpunct}
{\mcitedefaultendpunct}{\mcitedefaultseppunct}\relax
\EndOfBibitem
\bibitem[Bush \latin{et~al.}(2020)Bush, Volta, and Martin]{bush_diode}
Bush,~S.~N.; Volta,~T.~T.; Martin,~C.~R. Chemical sensing and chemoresponsive
  pumping with conical-pore polymeric membranes. \emph{Nanomaterials}
  \textbf{2020}, \emph{10}, 571\relax
\mciteBstWouldAddEndPuncttrue
\mciteSetBstMidEndSepPunct{\mcitedefaultmidpunct}
{\mcitedefaultendpunct}{\mcitedefaultseppunct}\relax
\EndOfBibitem
\bibitem[Xiong \latin{et~al.}(2019)Xiong, Zhang, Jiang, Yu, and Mao]{Xiong2019}
Xiong,~T.; Zhang,~K.; Jiang,~Y.; Yu,~P.; Mao,~L. {Ion current rectification:
  from nanoscale to microscale}. \emph{Science China Chemistry} \textbf{2019},
  \emph{62}, 1346--1359\relax
\mciteBstWouldAddEndPuncttrue
\mciteSetBstMidEndSepPunct{\mcitedefaultmidpunct}
{\mcitedefaultendpunct}{\mcitedefaultseppunct}\relax
\EndOfBibitem
\bibitem[Woermann(2003)]{woermann1}
Woermann,~D. Electrochemical transport properties of a cone-shaped nanopore:
  high and low electrical conductivity states depending on the sign of an
  applied electrical potential difference. \emph{Physical Chemistry Chemical
  Physics} \textbf{2003}, \emph{5}, 1853--1858\relax
\mciteBstWouldAddEndPuncttrue
\mciteSetBstMidEndSepPunct{\mcitedefaultmidpunct}
{\mcitedefaultendpunct}{\mcitedefaultseppunct}\relax
\EndOfBibitem
\bibitem[Woermann(2004)]{woermann2}
Woermann,~D. Electrochemical transport properties of a cone-shaped nanopore:
  revisited. \emph{Physical Chemistry Chemical Physics} \textbf{2004},
  \emph{6}, 3130--3132\relax
\mciteBstWouldAddEndPuncttrue
\mciteSetBstMidEndSepPunct{\mcitedefaultmidpunct}
{\mcitedefaultendpunct}{\mcitedefaultseppunct}\relax
\EndOfBibitem
\bibitem[Ghosal \latin{et~al.}(2019)Ghosal, Sherwood, and Chang]{conereview}
Ghosal,~S.; Sherwood,~J.~D.; Chang,~H.-C. Solid-state nanopore hydrodynamics
  and transport. \emph{Biomicrofluidics} \textbf{2019}, \emph{13}, 011301\relax
\mciteBstWouldAddEndPuncttrue
\mciteSetBstMidEndSepPunct{\mcitedefaultmidpunct}
{\mcitedefaultendpunct}{\mcitedefaultseppunct}\relax
\EndOfBibitem
\bibitem[Siwy(2006)]{Siwy2006}
Siwy,~Z.~S. {Ion-Current Rectification in Nanopores and Nanotubes with Broken
  Symmetry}. \emph{Advanced Functional Materials} \textbf{2006}, \emph{16},
  735--746\relax
\mciteBstWouldAddEndPuncttrue
\mciteSetBstMidEndSepPunct{\mcitedefaultmidpunct}
{\mcitedefaultendpunct}{\mcitedefaultseppunct}\relax
\EndOfBibitem
\bibitem[Boon \latin{et~al.}(2022)Boon, Veenstra, Dijkstra, and van Roij]{POF}
Boon,~W.~Q.; Veenstra,~T.~E.; Dijkstra,~M.; van Roij,~R. Pressure-sensitive ion
  conduction in a conical channel: Optimal pressure and geometry. \emph{Physics
  of Fluids} \textbf{2022}, \emph{34}, 101701\relax
\mciteBstWouldAddEndPuncttrue
\mciteSetBstMidEndSepPunct{\mcitedefaultmidpunct}
{\mcitedefaultendpunct}{\mcitedefaultseppunct}\relax
\EndOfBibitem
\bibitem[Poggioli \latin{et~al.}(2019)Poggioli, Siria, and
  Bocquet]{Poggioli2019}
Poggioli,~A.~R.; Siria,~A.; Bocquet,~L. {Beyond the Tradeoff: Dynamic
  Selectivity in Ionic Transport and Current Rectification}. \emph{The Journal
  of Physical Chemistry B} \textbf{2019}, \emph{123}, 1171--1185\relax
\mciteBstWouldAddEndPuncttrue
\mciteSetBstMidEndSepPunct{\mcitedefaultmidpunct}
{\mcitedefaultendpunct}{\mcitedefaultseppunct}\relax
\EndOfBibitem
\bibitem[Cheng and Guo(2007)Cheng, and Guo]{Cheng2007}
Cheng,~L.-J.; Guo,~L.~J. {Rectified Ion Transport through Concentration
  Gradient in Homogeneous Silica Nanochannels}. \emph{Nano Letters}
  \textbf{2007}, \emph{7}, 3165--3171\relax
\mciteBstWouldAddEndPuncttrue
\mciteSetBstMidEndSepPunct{\mcitedefaultmidpunct}
{\mcitedefaultendpunct}{\mcitedefaultseppunct}\relax
\EndOfBibitem
\bibitem[Graf \latin{et~al.}(2019)Graf, Lihter, Unuchek, Sarathy, Leburton,
  Kis, and Radenovic]{Graf2019}
Graf,~M.; Lihter,~M.; Unuchek,~D.; Sarathy,~A.; Leburton,~J.-P.; Kis,~A.;
  Radenovic,~A. {Light-Enhanced Blue Energy Generation Using MoS2 Nanopores}.
  \emph{Joule} \textbf{2019}, \emph{3}, 1549--1564\relax
\mciteBstWouldAddEndPuncttrue
\mciteSetBstMidEndSepPunct{\mcitedefaultmidpunct}
{\mcitedefaultendpunct}{\mcitedefaultseppunct}\relax
\EndOfBibitem
\bibitem[Vlassiouk \latin{et~al.}(2008)Vlassiouk, Smirnov, and
  Siwy]{siwy_saturation}
Vlassiouk,~I.; Smirnov,~S.; Siwy,~Z. Nanofluidic ionic diodes. Comparison of
  analytical and numerical solutions. \emph{Acs Nano} \textbf{2008}, \emph{2},
  1589--1602\relax
\mciteBstWouldAddEndPuncttrue
\mciteSetBstMidEndSepPunct{\mcitedefaultmidpunct}
{\mcitedefaultendpunct}{\mcitedefaultseppunct}\relax
\EndOfBibitem
\bibitem[{Dal Cengio} and Pagonabarraga(2019){Dal Cengio}, and
  Pagonabarraga]{Cengio2019}
{Dal Cengio},~S.; Pagonabarraga,~I. {Confinement-controlled rectification in a
  geometric nanofluidic diode}. \emph{The Journal of Chemical Physics}
  \textbf{2019}, \emph{151}, 044707\relax
\mciteBstWouldAddEndPuncttrue
\mciteSetBstMidEndSepPunct{\mcitedefaultmidpunct}
{\mcitedefaultendpunct}{\mcitedefaultseppunct}\relax
\EndOfBibitem
\bibitem[Kovarik \latin{et~al.}(2009)Kovarik, Zhou, and Jacobson]{rect1largeR}
Kovarik,~M.~L.; Zhou,~K.; Jacobson,~S.~C. Effect of conical nanopore diameter
  on ion current rectification. \emph{The Journal of Physical Chemistry B}
  \textbf{2009}, \emph{113}, 15960--15966\relax
\mciteBstWouldAddEndPuncttrue
\mciteSetBstMidEndSepPunct{\mcitedefaultmidpunct}
{\mcitedefaultendpunct}{\mcitedefaultseppunct}\relax
\EndOfBibitem
\bibitem[Lin \latin{et~al.}(2018)Lin, Yeh, and Siwy]{Lin2018}
Lin,~C.-y.; Yeh,~L.-h.; Siwy,~Z.~S. {Voltage-Induced Modulation of Ionic
  Concentrations and Ion Current Rectification in Mesopores with Highly Charged
  Pore Walls}. \emph{The Journal of Physical Chemistry Letters} \textbf{2018},
  \emph{9}, 393--398\relax
\mciteBstWouldAddEndPuncttrue
\mciteSetBstMidEndSepPunct{\mcitedefaultmidpunct}
{\mcitedefaultendpunct}{\mcitedefaultseppunct}\relax
\EndOfBibitem
\bibitem[Cervera \latin{et~al.}(2006)Cervera, Schiedt, Neumann, Maf{\'e}, and
  Ram{\'\i}rez]{cervera_exp_micropores_and_saturation}
Cervera,~J.; Schiedt,~B.; Neumann,~R.; Maf{\'e},~S.; Ram{\'\i}rez,~P. Ionic
  conduction, rectification, and selectivity in single conical nanopores.
  \emph{The Journal of chemical physics} \textbf{2006}, \emph{124},
  104706\relax
\mciteBstWouldAddEndPuncttrue
\mciteSetBstMidEndSepPunct{\mcitedefaultmidpunct}
{\mcitedefaultendpunct}{\mcitedefaultseppunct}\relax
\EndOfBibitem
\bibitem[Zhou \latin{et~al.}(2011)Zhou, Perry, and
  Jacobson]{zhou_exp_micropores}
Zhou,~K.; Perry,~J.~M.; Jacobson,~S.~C. Transport and sensing in nanofluidic
  devices. \emph{Annual Review of Analytical Chemistry} \textbf{2011},
  \emph{4}, 321--341\relax
\mciteBstWouldAddEndPuncttrue
\mciteSetBstMidEndSepPunct{\mcitedefaultmidpunct}
{\mcitedefaultendpunct}{\mcitedefaultseppunct}\relax
\EndOfBibitem
\bibitem[Jubin \latin{et~al.}(2018)Jubin, Poggioli, Siria, and
  Bocquet]{Jubin2018}
Jubin,~L.; Poggioli,~A.; Siria,~A.; Bocquet,~L. {Dramatic pressure-sensitive
  ion conduction in conical nanopores}. \emph{Proceedings of the National
  Academy of Sciences of the United States of America} \textbf{2018},
  \emph{115}, 4063--4068\relax
\mciteBstWouldAddEndPuncttrue
\mciteSetBstMidEndSepPunct{\mcitedefaultmidpunct}
{\mcitedefaultendpunct}{\mcitedefaultseppunct}\relax
\EndOfBibitem
\bibitem[Lin \latin{et~al.}(2019)Lin, Combs, Su, Yeh, and Siwy]{Lin2019}
Lin,~C.-Y.; Combs,~C.; Su,~Y.-S.; Yeh,~L.-H.; Siwy,~Z.~S. {Rectification of
  Concentration Polarization in Mesopores Leads To High Conductance Ionic
  Diodes and High Performance Osmotic Power}. \emph{Journal of the American
  Chemical Society} \textbf{2019}, \emph{141}, 3691--3698\relax
\mciteBstWouldAddEndPuncttrue
\mciteSetBstMidEndSepPunct{\mcitedefaultmidpunct}
{\mcitedefaultendpunct}{\mcitedefaultseppunct}\relax
\EndOfBibitem
\bibitem[He \latin{et~al.}(2017)He, Zhang, Li, Jiang, Yu, and Mao]{He2017}
He,~X.; Zhang,~K.; Li,~T.; Jiang,~Y.; Yu,~P.; Mao,~L. {Micrometer-Scale Ion
  Current Rectification at Polyelectrolyte Brush-Modified Micropipets}.
  \emph{Journal of the American Chemical Society} \textbf{2017}, \emph{139},
  1396--1399\relax
\mciteBstWouldAddEndPuncttrue
\mciteSetBstMidEndSepPunct{\mcitedefaultmidpunct}
{\mcitedefaultendpunct}{\mcitedefaultseppunct}\relax
\EndOfBibitem
\bibitem[Lee \latin{et~al.}(2012)Lee, Joly, Siria, Biance, Fulcrand, and
  Bocquet]{Lee2012}
Lee,~C.; Joly,~L.; Siria,~A.; Biance,~A.-L.; Fulcrand,~R.; Bocquet,~L. {Large
  Apparent Electric Size of Solid-State Nanopores Due to Spatially Extended
  Surface Conduction}. \emph{Nano Letters} \textbf{2012}, \emph{12},
  4037--4044\relax
\mciteBstWouldAddEndPuncttrue
\mciteSetBstMidEndSepPunct{\mcitedefaultmidpunct}
{\mcitedefaultendpunct}{\mcitedefaultseppunct}\relax
\EndOfBibitem
\bibitem[Hall(1975)]{hall1975access}
Hall,~J.~E. Access resistance of a small circular pore. \emph{The Journal of
  general physiology} \textbf{1975}, \emph{66}, 531--532\relax
\mciteBstWouldAddEndPuncttrue
\mciteSetBstMidEndSepPunct{\mcitedefaultmidpunct}
{\mcitedefaultendpunct}{\mcitedefaultseppunct}\relax
\EndOfBibitem
\bibitem[Xiang \latin{et~al.}(2022)Xiang, Dong, Liang, Zhang, and
  Guan]{pyramidal}
Xiang,~F.; Dong,~M.; Liang,~S.; Zhang,~W.; Guan,~W. Modeling pyramidal silicon
  nanopores with effective ion transport. \emph{Nanotechnology} \textbf{2022},
  \relax
\mciteBstWouldAddEndPunctfalse
\mciteSetBstMidEndSepPunct{\mcitedefaultmidpunct}
{}{\mcitedefaultseppunct}\relax
\EndOfBibitem
\bibitem[Yossifon \latin{et~al.}(2010)Yossifon, Mushenheim, Chang, and
  Chang]{yossifon}
Yossifon,~G.; Mushenheim,~P.; Chang,~Y.-C.; Chang,~H.-C. Eliminating the
  limiting-current phenomenon by geometric field focusing into nanopores and
  nanoslots. \emph{Physical Review E} \textbf{2010}, \emph{81}, 046301\relax
\mciteBstWouldAddEndPuncttrue
\mciteSetBstMidEndSepPunct{\mcitedefaultmidpunct}
{\mcitedefaultendpunct}{\mcitedefaultseppunct}\relax
\EndOfBibitem
\bibitem[Ma \latin{et~al.}(2018)Ma, Guo, Jia, and Xie]{Ma2018}
Ma,~Y.; Guo,~J.; Jia,~L.; Xie,~Y. {Entrance Effects Induced Rectified Ionic
  Transport in a Nanopore/Channel}. \emph{ACS Sensors} \textbf{2018}, \emph{3},
  167--173\relax
\mciteBstWouldAddEndPuncttrue
\mciteSetBstMidEndSepPunct{\mcitedefaultmidpunct}
{\mcitedefaultendpunct}{\mcitedefaultseppunct}\relax
\EndOfBibitem
\bibitem[Su \latin{et~al.}(2018)Su, Ji, Tang, Li, Feng, Cao, Jiang, and
  Guo]{Su2018}
Su,~J.; Ji,~D.; Tang,~J.; Li,~H.; Feng,~Y.; Cao,~L.; Jiang,~L.; Guo,~W.
  {Anomalous Pore-Density Dependence in Nanofluidic Osmotic Power Generation}.
  \emph{Chinese Journal of Chemistry} \textbf{2018}, \emph{36}, 417--420\relax
\mciteBstWouldAddEndPuncttrue
\mciteSetBstMidEndSepPunct{\mcitedefaultmidpunct}
{\mcitedefaultendpunct}{\mcitedefaultseppunct}\relax
\EndOfBibitem
\bibitem[Kowalczyk \latin{et~al.}(2011)Kowalczyk, Grosberg, Rabin, and
  Dekker]{dekker_resistance}
Kowalczyk,~S.~W.; Grosberg,~A.~Y.; Rabin,~Y.; Dekker,~C. Modeling the
  conductance and DNA blockade of solid-state nanopores. \emph{Nanotechnology}
  \textbf{2011}, \emph{22}, 315101\relax
\mciteBstWouldAddEndPuncttrue
\mciteSetBstMidEndSepPunct{\mcitedefaultmidpunct}
{\mcitedefaultendpunct}{\mcitedefaultseppunct}\relax
\EndOfBibitem
\bibitem[Kirby(2010)]{kirby_micro}
Kirby,~B.~J. \emph{Micro-and nanoscale fluid mechanics: transport in
  microfluidic devices}; Cambridge university press, 2010\relax
\mciteBstWouldAddEndPuncttrue
\mciteSetBstMidEndSepPunct{\mcitedefaultmidpunct}
{\mcitedefaultendpunct}{\mcitedefaultseppunct}\relax
\EndOfBibitem
\bibitem[Delgado \latin{et~al.}(2005)Delgado, Gonz{\'a}lez-Caballero, Hunter,
  Koopal, and Lyklema]{IUPAC}
Delgado,~A.~V.; Gonz{\'a}lez-Caballero,~F.; Hunter,~R.; Koopal,~L.~K.;
  Lyklema,~J. Measurement and interpretation of electrokinetic phenomena (IUPAC
  technical report). \emph{Pure and Applied Chemistry} \textbf{2005},
  \emph{77}, 1753--1805\relax
\mciteBstWouldAddEndPuncttrue
\mciteSetBstMidEndSepPunct{\mcitedefaultmidpunct}
{\mcitedefaultendpunct}{\mcitedefaultseppunct}\relax
\EndOfBibitem
\bibitem[Harned and Nuttall(1949)Harned, and Nuttall]{difcon1}
Harned,~H.~S.; Nuttall,~R.~L. The differential diffusion coefficient of
  potassium chloride in aqueous solutions. \emph{Journal of the American
  Chemical Society} \textbf{1949}, \emph{71}, 1460--1463\relax
\mciteBstWouldAddEndPuncttrue
\mciteSetBstMidEndSepPunct{\mcitedefaultmidpunct}
{\mcitedefaultendpunct}{\mcitedefaultseppunct}\relax
\EndOfBibitem
\bibitem[dif(2004)]{difcon2}
\emph{CRC handbook of chemistry and physics}; CRC press, 2004; Vol.~85\relax
\mciteBstWouldAddEndPuncttrue
\mciteSetBstMidEndSepPunct{\mcitedefaultmidpunct}
{\mcitedefaultendpunct}{\mcitedefaultseppunct}\relax
\EndOfBibitem
\bibitem[Adamson and Gast(1967)Adamson, and Gast]{physicalchem}
Adamson,~A.~W.; Gast,~A.~P. \emph{Physical Chemistry of Surfaces}; Interscience
  publishers New York, 1967; Vol. 150\relax
\mciteBstWouldAddEndPuncttrue
\mciteSetBstMidEndSepPunct{\mcitedefaultmidpunct}
{\mcitedefaultendpunct}{\mcitedefaultseppunct}\relax
\EndOfBibitem
\bibitem[Szekeres \latin{et~al.}(1998)Szekeres, D{\'e}k{\'a}ny, and
  De~Keizer]{surfpotKCl1}
Szekeres,~M.; D{\'e}k{\'a}ny,~I.; De~Keizer,~A. Adsorption of dodecyl
  pyridinium chloride on monodisperse porous silica. \emph{Colloids and
  Surfaces A: Physicochemical and Engineering Aspects} \textbf{1998},
  \emph{141}, 327--336\relax
\mciteBstWouldAddEndPuncttrue
\mciteSetBstMidEndSepPunct{\mcitedefaultmidpunct}
{\mcitedefaultendpunct}{\mcitedefaultseppunct}\relax
\EndOfBibitem
\bibitem[Janusz(1996)]{surfpotKCl2}
Janusz,~W. The Structure of the Electrical Double Layer at the LiChrospher-Type
  Adsorbent/Aqueous Electrolyte Solution Interface. \emph{Adsorption Science \&
  Technology} \textbf{1996}, \emph{14}, 151--161\relax
\mciteBstWouldAddEndPuncttrue
\mciteSetBstMidEndSepPunct{\mcitedefaultmidpunct}
{\mcitedefaultendpunct}{\mcitedefaultseppunct}\relax
\EndOfBibitem
\bibitem[Werkhoven and van Roij(2020)Werkhoven, and van
  Roij]{werkhovensoftmatter}
Werkhoven,~B.~L.; van Roij,~R. Coupled water, charge and salt transport in
  heterogeneous nano-fluidic systems. \emph{Soft Matter} \textbf{2020},
  \emph{16}, 1527--1537\relax
\mciteBstWouldAddEndPuncttrue
\mciteSetBstMidEndSepPunct{\mcitedefaultmidpunct}
{\mcitedefaultendpunct}{\mcitedefaultseppunct}\relax
\EndOfBibitem
\bibitem[Zhao \latin{et~al.}(2015)Zhao, Ebeling, Siretanu, van~den Ende, and
  Mugele]{regulation}
Zhao,~C.; Ebeling,~D.; Siretanu,~I.; van~den Ende,~D.; Mugele,~F. Extracting
  local surface charges and charge regulation behavior from atomic force
  microscopy measurements at heterogeneous solid-electrolyte interfaces.
  \emph{Nanoscale} \textbf{2015}, \emph{7}, 16298--16311\relax
\mciteBstWouldAddEndPuncttrue
\mciteSetBstMidEndSepPunct{\mcitedefaultmidpunct}
{\mcitedefaultendpunct}{\mcitedefaultseppunct}\relax
\EndOfBibitem
\bibitem[Iler(1955)]{silica1}
Iler,~R.~K. \emph{The colloid chemistry of silica and silicates}; LWW, 1955;
  Vol.~80; p 666\relax
\mciteBstWouldAddEndPuncttrue
\mciteSetBstMidEndSepPunct{\mcitedefaultmidpunct}
{\mcitedefaultendpunct}{\mcitedefaultseppunct}\relax
\EndOfBibitem
\bibitem[Kosmulski(2001)]{kosmulski}
Kosmulski,~M. \emph{Chemical properties of material surfaces}; CRC press, 2001;
  Vol. 102\relax
\mciteBstWouldAddEndPuncttrue
\mciteSetBstMidEndSepPunct{\mcitedefaultmidpunct}
{\mcitedefaultendpunct}{\mcitedefaultseppunct}\relax
\EndOfBibitem
\bibitem[White and Bund(2008)White, and Bund]{rect1}
White,~H.~S.; Bund,~A. Ion current rectification at nanopores in glass
  membranes. \emph{Langmuir} \textbf{2008}, \emph{24}, 2212--2218\relax
\mciteBstWouldAddEndPuncttrue
\mciteSetBstMidEndSepPunct{\mcitedefaultmidpunct}
{\mcitedefaultendpunct}{\mcitedefaultseppunct}\relax
\EndOfBibitem
\bibitem[Wen \latin{et~al.}(2019)Wen, Zeng, Li, Zhang, and Zhang]{rect2}
Wen,~C.; Zeng,~S.; Li,~S.; Zhang,~Z.; Zhang,~S.-L. On rectification of ionic
  current in nanopores. \emph{Analytical chemistry} \textbf{2019}, \emph{91},
  14597--14604\relax
\mciteBstWouldAddEndPuncttrue
\mciteSetBstMidEndSepPunct{\mcitedefaultmidpunct}
{\mcitedefaultendpunct}{\mcitedefaultseppunct}\relax
\EndOfBibitem
\bibitem[Proctor(2021)]{rect3}
Proctor,~J.~E. Theory of Ion Transport and Ion Current Rectification in
  Nanofluidic Diodes. \textbf{2021}, 105\relax
\mciteBstWouldAddEndPuncttrue
\mciteSetBstMidEndSepPunct{\mcitedefaultmidpunct}
{\mcitedefaultendpunct}{\mcitedefaultseppunct}\relax
\EndOfBibitem
\bibitem[Inamdar \latin{et~al.}(2009)Inamdar, Bhat, and Haram]{Inamdar2009}
Inamdar,~S.~N.; Bhat,~M.~A.; Haram,~S.~K. {Construction of Ag/AgCl Reference
  Electrode from Used Felt-Tipped Pen Barrel for Undergraduate Laboratory}.
  \emph{Journal of Chemical Education} \textbf{2009}, \emph{86}, 355\relax
\mciteBstWouldAddEndPuncttrue
\mciteSetBstMidEndSepPunct{\mcitedefaultmidpunct}
{\mcitedefaultendpunct}{\mcitedefaultseppunct}\relax
\EndOfBibitem
\bibitem[Landau and Lifshitz(2013)Landau, and Lifshitz]{landau}
Landau,~L.~D.; Lifshitz,~E.~M. \emph{Fluid Mechanics: Landau and Lifshitz:
  Course of Theoretical Physics, Volume 6}; Elsevier, 2013; Vol.~6\relax
\mciteBstWouldAddEndPuncttrue
\mciteSetBstMidEndSepPunct{\mcitedefaultmidpunct}
{\mcitedefaultendpunct}{\mcitedefaultseppunct}\relax
\EndOfBibitem
\end{mcitethebibliography}

\providecommand{\latin}[1]{#1}
\makeatletter
\providecommand{\doi}
  {\begingroup\let\do\@makeother\dospecials
  \catcode`\{=1 \catcode`\}=2 \doi@aux}
\providecommand{\doi@aux}[1]{\endgroup\texttt{#1}}
\makeatother
\providecommand*\mcitethebibliography{\thebibliography}
\csname @ifundefined\endcsname{endmcitethebibliography}
  {\let\endmcitethebibliography\endthebibliography}{}

\newpage
\section{Supplementary}
\beginsupplement

\subsection{Supplementary Information 1: Atomic Force Microscopy profile of tapered feature made with Focused Ion Beam}

\begin{figure}[h]
	\includegraphics[width=.6\textwidth]{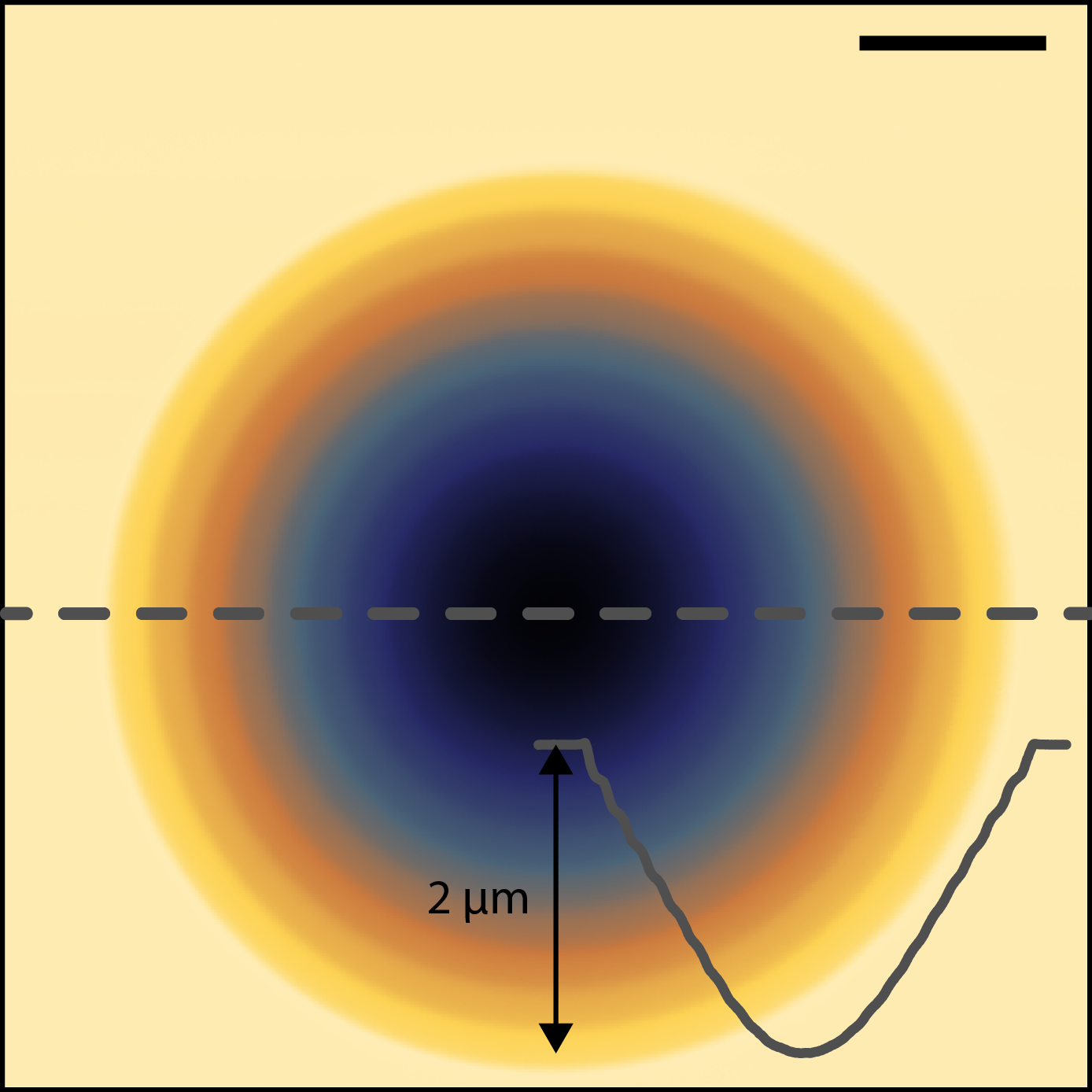}
	\caption{AFM image of a tapered hole with 5 $\mu$m radius made with FIB at the edge of the membrane on top of the supporting frame. The centerline profile (grey dashed line) is depicted in the image (grey solid line), with the total depth of the hole being 2 $\mu$m. The colour scale runs from light yellow (high) to dark blue (low), the scale bar is 2 $\mu$m}
	\label{SI-1}
\end{figure}

Atomic Force Microscopy (AFM) image of a profile resulting from our Focused Ion Beam milling (FIB) protocol writing concentric circles, shown for a hole with a 5 $\mu$m base radius. Milling is done at the edge of the membrane where it is supported by a silicon frame, allowing AFM imaging. The image and inset profile illustrate the smooth profile. The depth of the hole is 2 $\mu$m. This depth appears to be set by the underlying frame as the milling rate was found to be much slower once the frame is reached. 

\newpage

\subsection{Supplementary Information 2: Illustration of the experimental setup}

\begin{figure}[h]
	\includegraphics[width=.4\textwidth]{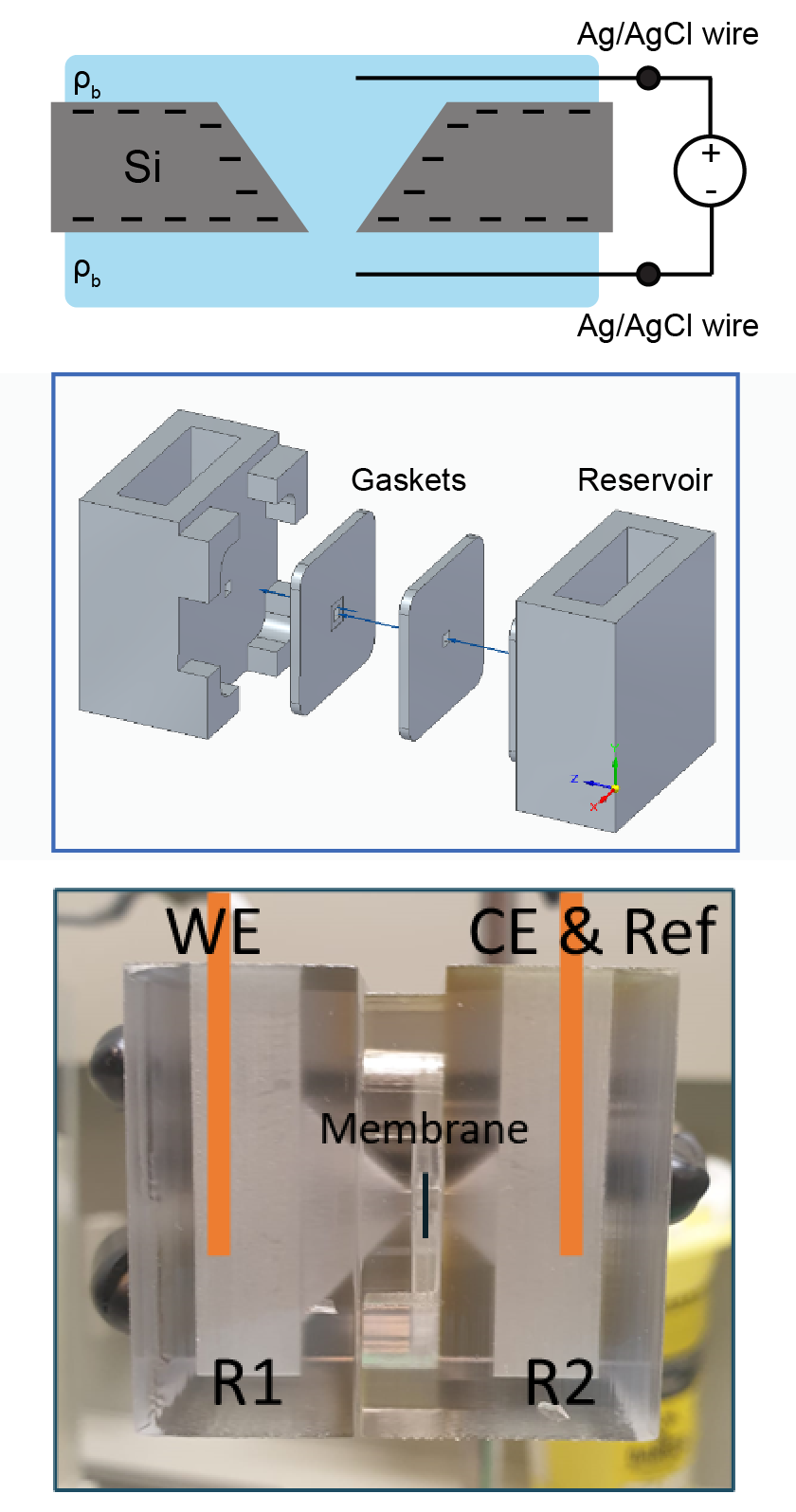}
	\caption{\textbf{(top)} Schematic of the experimental setup, with 2 reservoirs of equal concentration $\rho_{\rm{b}}$ of KCl on either side of the membrane. \textbf{(middle)} 3D drawing of the setup, consisting of 2 electrolyte reservoirs and 2 PDMS gaskets. \textbf{(bottom)} Photo of the experimental setup with the membrane and electrode positions indicated. The working electrode (WE) is connected to a Ag/AgCl wire placed in reservoir R1 facing the large opening of the pore, and the counter and reference electrodes (CE and ref) are both connected to the second Ag/AgCl wire in reservoir R2, facing the small pore opening. All components are clamped together using a laboratory clamp.}
	\label{SI-2}
\end{figure}

Schematics and photo of the experimental setup. The setup consists of 2 electrolyte reservoirs (3D printed in a transparent commercial polymer \textit{VeroClear}) and 2 Polydimethylsiloxane (PDMS) gaskets. The reservoirs have a tapered opening leading towards the membrane, to facilitate filling of the channel without trapping air. The PDMS gaskets (thickness $\approx$ 1 mm) were cast in an aluminium mould to obtain the hole in the center, and a recess to fit the membrane. Measurements are done in a 2-electrode configuration with the working electrode (WE) connected to a Ag/AgCl wire in the reservoir facing the large pore opening (R1) and the counter,- and reference electrode (CE and Ref) connected to a Ag/AgCl wire in the reservoir facing the small pore opening (R2), as indicated in the photo. The membrane is placed between the gaskets and the setup is tightly squeezed and held together by a laboratory clamp.

\newpage

\subsection{Supplementary Information 3: Current-voltage characterization of as-received membrane}

\begin{figure}[h]
	\includegraphics[width=.6\textwidth]{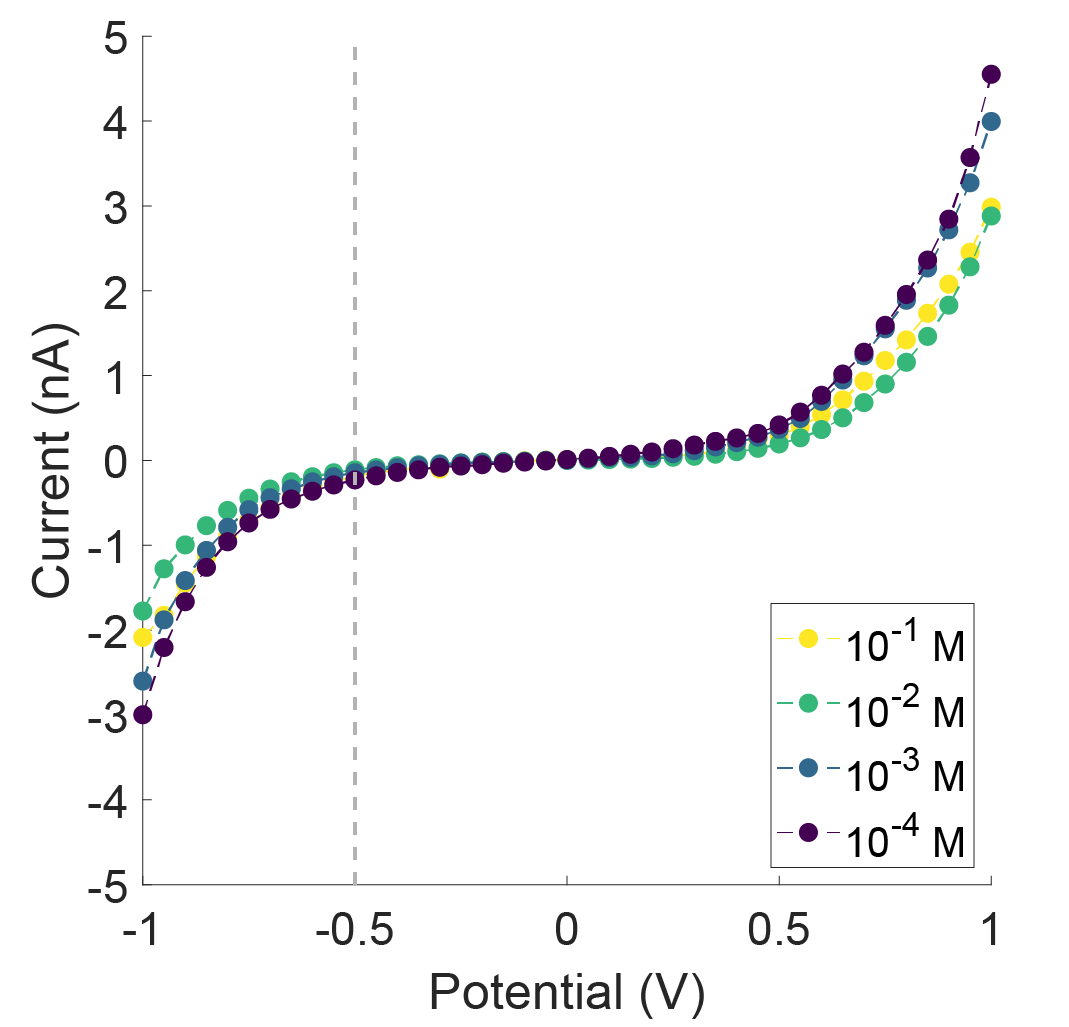}
	\caption{I-V reference curves determining the leakage current of an as-received membrane, without any pore. The leakage current is the same for different concentrations of KCl, as indicated by the legend. The grey dashed line at -0.5 V indicates the potential at which data in Fig. (\ref{Fig2}) in the main text is normalized.}
	\label{SI-3}
\end{figure}

Current-voltage (I-V) characterization of an as-received membrane for different concentrations of KCl. However, the leakage current becomes significant at potentials $|\Delta\psi|\geq$ 0.5 V, in particular considering our measurements at low concentrations where the total current through the pore is in the 1-10 nA range. The leakage current does not show a clear concentration dependence, despite differing slightly between different measurements. It is worth mentioning that the setup is disassembled for filling with a different concentration, illustrating the robustness of the sealing by the gaskets for subsequent measurements at the potentials considered in the main text. The average of these measurements is used as lower bound of the shaded region in Figure \ref{Fig2}(a-d) in the main text to illustrate the potential contribution by the leakage current, capped at $I$ = 0 for 2 of the curves in Figure \ref{Fig2}(c). 

\newpage

\subsection{Supplementary Information 4: Calculation of the electric electric field}

In this section we will calculate the electric field $-\nabla\psi$ not only within the pore but also in both reservoirs. This analysis supposes that the space charge $\rho_{\rm{e}}$ outside the EDL is negligible, ensuring that the electric field is divergence free $\nabla^2\psi=0$. We assume (i) that the electric field in the bulk reservoirs far from the pore $\sqrt{r^2+x^2}\gg R_{\rm{b}}$ is isotropic and decays like an electric monopole by the inverse square law $\propto 1/(r^2+x^2)$ and, (ii) that no electric field permeates the channel walls. The far-from-pore assumption (i) solution breaks down in the near-pore region $r^2+x^2\simeq R_{\rm{b}}^2$ where the electric field diverges and therefore a characteristic cut-off length scale for this asymptotic decay has to be identified. Natural length scales would be the tip and base radii $R_{\rm{t}}$ and $R_{\rm{b}}$ near which the far-from-pore solution fails, but to obtain quantitative agreement with numerical calculations we have to multiply the base and tip radii by $\pi/4$. Choosing this factor will also reproduce the exact resistance for a cylindrical, 2D-pore as derived by Hall\cite{hall1975access}. Following assumption (ii) the field inside the conical pore $0<x<L$ scales as $\partial_x\psi\propto1/(\pi R^{2}(x))$ as the total, radially integrated, lateral electric field has to be constant. Combining these expression we find the electric field over the centre axis $r=0$ is given by
\begin{equation}
    -\partial_x\psi(x,r=0)=
    \begin{dcases}
    \frac{\alpha\pi^2}{4} \Delta\psi\bigg(\frac{\pi}{4} R_{\rm{b}}-x\bigg)^{-2}  & \text{if} \ x \textless 0;\\
    \alpha \Delta\psi R^{-2}(x) & \text{if} \ 0\textless x \textless L; \\
    \frac{\alpha\pi^2}{4} \Delta\psi \bigg(x+\frac{\pi}{4}  R_{\rm{t}}-L\bigg)^{-2}  & \text{if} \ x \textgreater L, \tag{S1}
    \end{dcases}
     \label{E_field}
\end{equation}
\begin{figure}[t!]
	\includegraphics[width=.45\textwidth]{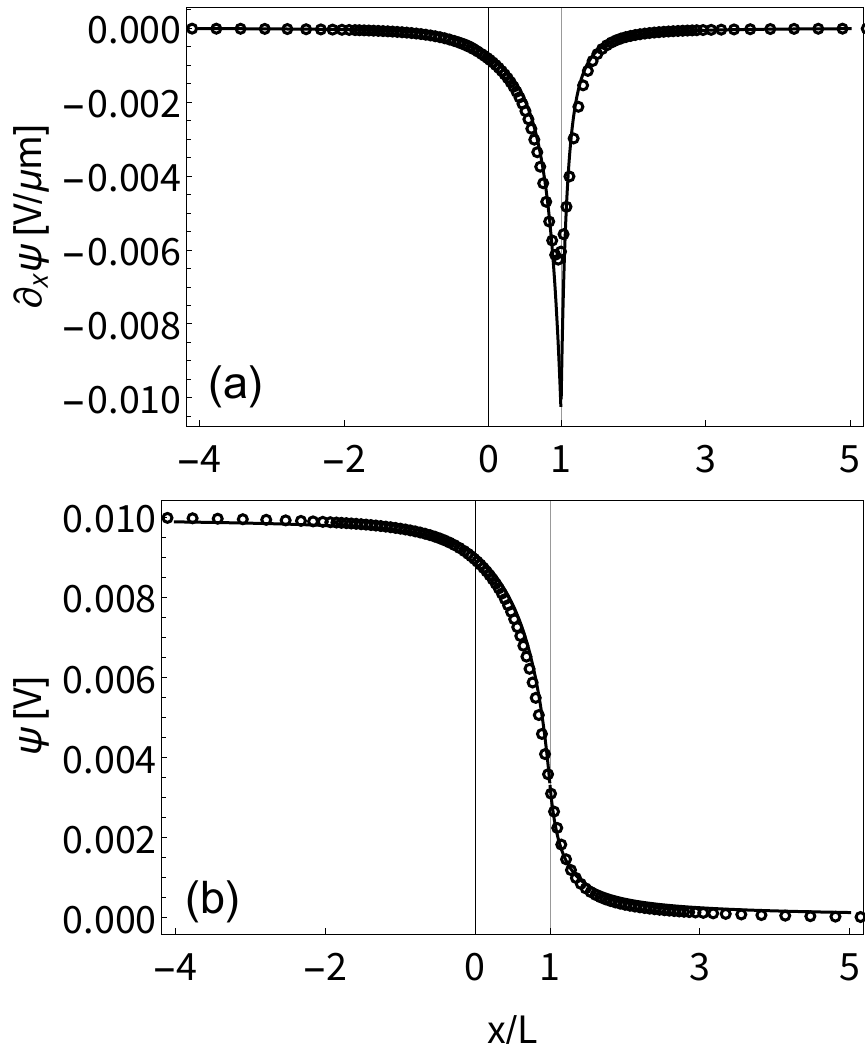}
	\caption{(a) Gradient of the electric potential $\partial_x\psi(x)$ and (b) electric potential $\psi(x)$ along the central axis $r=0$ for the T1 geometry (see the main text) at a vanishing surface potential and $\rho_{\rm{b}}=1$ mM for  $\Delta\psi=0.01$ V with symbols from numerical calculations and lines resulting from Eq.(\hyref{E_field}). This parameter set is representative for our experimental system at high concentrations, where surface conductance is negligible. The base and tip locations are at $x=0$ and $x=L$ denoted by vertical lines. There is good agreement between numerical and analytic results.}
	\label{FigE}
\end{figure}
where the constant length $\alpha=R_{\rm{b}} R_{\rm{t}}/(4L+\pi (R_{\rm{b}}+ R_{\rm{t}}))$ can be found by requiring that the electric field is continuous at the pore edges and the total potential drop equals $\psi(-\infty)-\psi(\infty)=\Delta\psi$ and the maximum electric field (at the tip) is equal to $\alpha\Delta\psi/R_{\rm{t}}^2$. We have chosen to evaluate the electric field on the center line where the field is purely axial as to give an explicit expression for one of the vector components of $-\nabla\psi$. Furthermore we note that this component is of greatest interest as it is responsible for the axial currents through the pore. In Fig.\hyref{FigE} we compare the analytic expression of the electric field (Eq.\hyref{E_field}) over the central axes and find good agreement with numerical results.

\newpage
\clearpage

\subsection{Supplementary Information 5: Scanning Electron Microscope image of pore T1, after measurements}

\begin{figure}[h]
	\includegraphics[width=.6\textwidth]{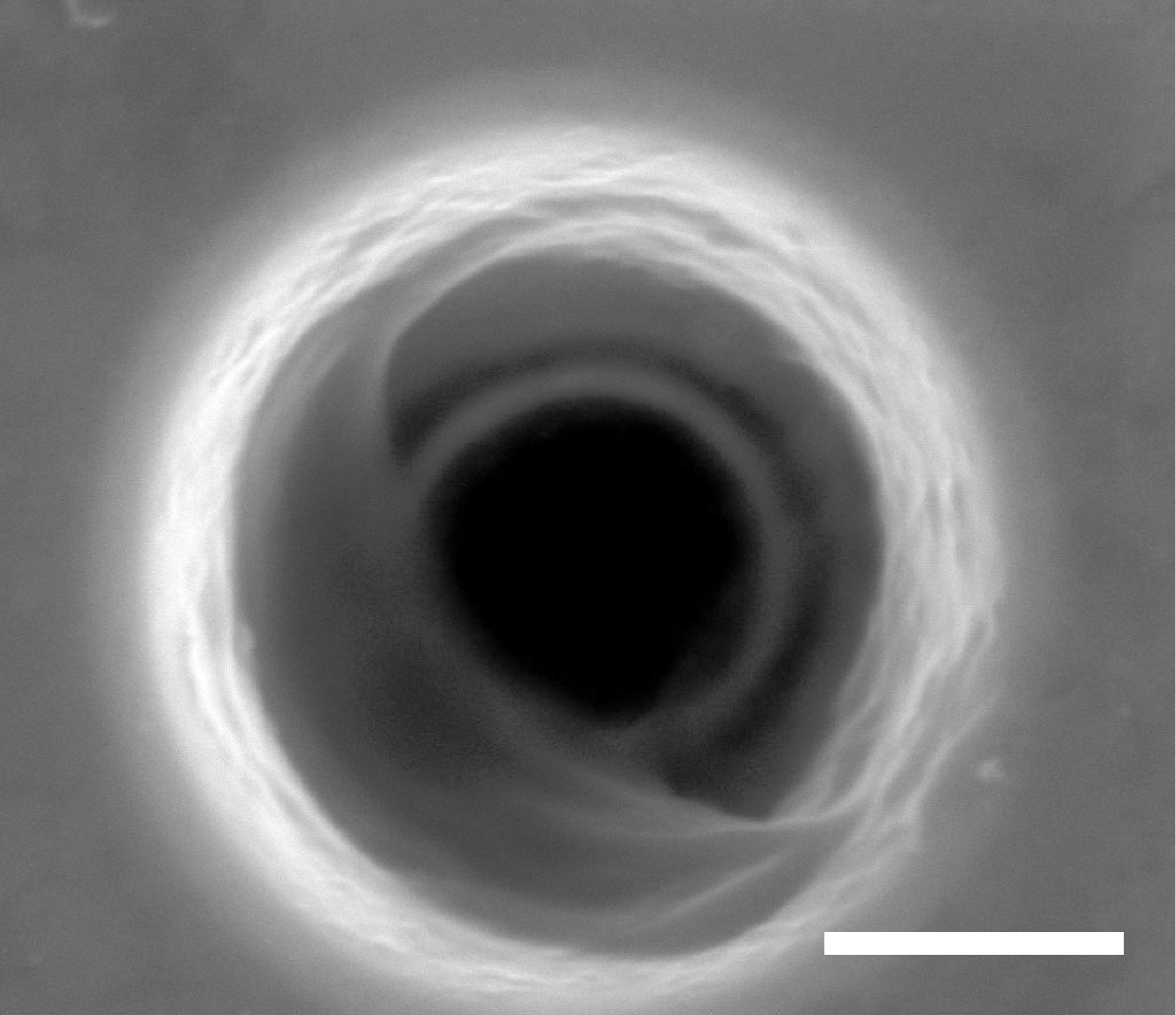}
	\caption{SEM image of the tapered pore T1 after the measurements. The scale bar is 1 $\mu$m}
	\label{SI-6}
\end{figure}

Fig.\hyref{SI-6} shows a scanning electron microscope (SEM) image of the tapered pore T1 in the main text, after the measurements. Both partial clogging and roughening of the pore are observed thereby changing the pore geometry. Hence clogging will change the pore conductance, which likely explains the variation of conductance over a series of experiments. The chronology of experiments is from low to high-concentration.
\newpage

\subsection{Supplementary Information 6: Inlet-outlet concentration polarization}
In this section we will construct a far-from-pore solution demonstrating concentration-polarization in the bulk reservoir with $x\textless 0$ connected to the pore base. Instead of the cylindrical $(x,r,\theta)$ coordinates used in the main text it will be convenient to treat the problem in spherical $(s, \phi, \theta)$ coordinates with $s^2=x^2+r^2$ and $\cos\phi=x/\sqrt{x^2+r^2}$. We consider fluxes far from the pore opening $s\gg L$ where the electric field Eq.(\hyref{E_field}) simplifies to $\partial_s\psi\simeq-\alpha \pi^2 \Delta\psi/(4s^2$). The aim here is to calculate $\hat{\rho}_{\rm{s}}(s)$, with $\hat{\cdots}=(2\pi s^2)^{-1}\int_0^{2\pi}\int_{\pi/2}^{\pi} \cdots s^2\sin(\phi)\dd \phi \dd\theta$ the average over a hemisphere centered on the origin extending in the bulk with radius $s$. The hemispherical average $\hat{\rho}_{\rm{s}}(s)$ is not representative for the local concentration $\rho_{\rm{s}}(s,\theta,\phi)$ which is expected to have a large $\phi$ dependence as the electric double layer is localized at $\phi=\pi/2$ and the far-from-pore Landau-Squire\cite{landau} solution for the fluid flow $\mathbf{u}(s,\phi)$ is much larger at $\phi=\pi$ than at $\phi=\pi/2$. Both these complicating factors will expectedly yield a concentration profile with larger deviations from bulk concentration near the membrane surface $\phi=\pi/2$ compared to $\phi=\pi$. Nevertheless our expression for $\hat{\rho}_{\rm{s}}(s)$ can explain two experimental observations, (i) concentration-polarization in the bulk reservoir is expected in the small-pore limit $L/R_{\rm{b}}\simeq 1$ and (ii) the concentration profile extends long distances into the bulk, exhibiting long-ranged, inverse-square decay. \\

Integrating the radial component $j_{\rm{s},\it{s}}(s,\phi)$ of the salt flux to obtain the total salt flux $\hat{J}(s)=\int_0^{2\pi}\dd \theta \int_{\pi/2}^{\pi}\dd\phi \sin(\phi)s^2 j_{\rm{s},\it{s}}(s,\phi)$ and imposing the stationarity condition $\partial_s\hat{J}=0$ we find a differential equation for the concentration $\hat{\rho}_{\rm{s}}(s)$ averaged over a hemisphere,
\begin{equation}
   D(2\pi\partial_{s}(s^2 \partial_{s} \hat{\rho})-\frac{\pi ^3\alpha\sigma}{2s^2}  \frac{e\Delta\psi}{k_{\rm{B}}T})+Q\partial_{s}\hat{\rho}_{\rm{s}}=0,\tag{S2}
   \label{Jhat}
\end{equation}
 with $s$ denoting the radius of the hemisphere over which the concentration is averaged, $\alpha$ being defined below Eq.(\hyref{E_field}) and where $\int_0^{2\pi}\dd \theta \int_{\pi/2}^{\pi}\dd\phi \sin(\phi)s^2\rho_{\rm{e}}(s,\phi)=-2\pi s \sigma$ stems from hemispherical charge-neutrality. Furthermore we made the approximation that flow can be considered to be isotropic which combined with incompressibility yields $2\pi s^2u_s(s)=-Q$, where the minus sign was added so that a radially inward flow in the bulk reservoir results in a positive $Q$ following the convention in the main text. Solving Eq.(\hyref{Jhat}) for $\hat{\rho}_{\rm{s}}$ with bulk boundary conditions $\hat{\rho}_{\rm{s}}(\infty)=\partial_s\hat{\rho}_{\rm{s}}(\infty)=0$ we find
\begin{equation}
    \hat{\rho}_{\rm{s}}(s)-2\rho_{\rm{b}}= \Delta\rho_{\rm{res}} \bigg( \exp\big(\frac{l_{\rm{Pe}}}{s}\big)-\frac{l_{\rm{Pe}}}{s}-1\bigg)\stackrel{s\gg |l_{\rm{Pe}}|}{\simeq}  \frac{\Delta\rho_{\rm{res}}}{2}\bigg(\frac{l_{\rm{Pe}}}{s}\bigg)^2,\tag{S3}
    \label{res_conc}
\end{equation}
where the measure for the concentration profile extending into the reservoir is 
\begin{equation}
    \Delta\rho_{\rm{res}}=\frac{\pi\sigma}{4l_{\rm{Pe}}^2} \frac{e\Delta\psi}{k_{\rm{B}}T} \bigg[ \frac{R_{\rm{b}}R_{\rm{t}}}{4L/\pi+R_{\rm{b}}+R_{\rm{t}}}\bigg],\tag{4}
\end{equation}
with the P\'eclet length $l_{\rm{Pe}}=Q/2\pi D$ signifying the distance from the origin at which advective and diffusive transport rates are equal. Note that $Q$ and hence the P\'eclet length has a sign. The term in square brackets vanishes in the long-channel-limit as the electric-field in the bulk and correspondingly $\Delta\rho_{\rm{res}}$ go to zero in this limit, which shows that no pore-pore interactions are expected for long, thin pores. While our solution was specifically derived for the base reservoir with $x<0$ and $\phi\in[\pi/2,\pi]$, our solution Eq.(\hyref{res_conc}) is valid in the tip-connected reservoir with $x>L$ and $\phi\in[0,\pi/2]$ when interchanging $-\Delta\rho_{\rm{res}}$ for $\Delta\rho_{\rm{res}}$ and $-l_{\rm{Pe}}$ for $l_{\rm{Pe}}$ as the flow and electric field are anti-symmetric between tip- and base-connected reservoirs. Due to the anti-symmetry of the far-from-pore solutions the depletion in one reservoir leads to a compensating excess in the other reservoir for $s\gg  |l_{\rm{Pe}}|$ (where the asymptotic decay is independent from $l_{\rm{Pe}}$) and the only contribution to ICR is expected from the near-pore region. The unphysical divergence of the concentration profile near the pore for positive (inward) flows $s\ll l_{\rm{Pe}}$ prevents us from connecting the far-from-pore solution to the near-pore region. In this regime the large inward flow sweeps up the concentration profile and concentrates it in the near-pore region where the far-from-pore solution breaks down. \\

As the flow is always inwards for one of the two reservoirs there is no scenario where the far-from-pore solution can be used to describe the entire experimental system. We note that this focusing of the concentration profile near the base for positive flows also complicates numerical calculations: a significant effort was made to obtain numerical calculations from COMSOL, however no finite-element system could be created that was stable beyond a very narrow parameter regime. Our numerical calculations always showed reservoir concentration polarization in some form. The near-pore solution is expected to very sensitively depend on all experimental length scales, including P\'eclet and Dukhin length. A "holistic" model describing the entire concentration profile extending over both the reservoirs and pore would be desirable as it would allow for quantitative predictions without an "apparent" $\psi_0$ as fit parameter. This problem is left for future study.

\newpage

\subsection{Supplementary Information 7: Conductances at four different concentrations}

\begin{figure}[b!]
	\includegraphics[width=0.8\textwidth]{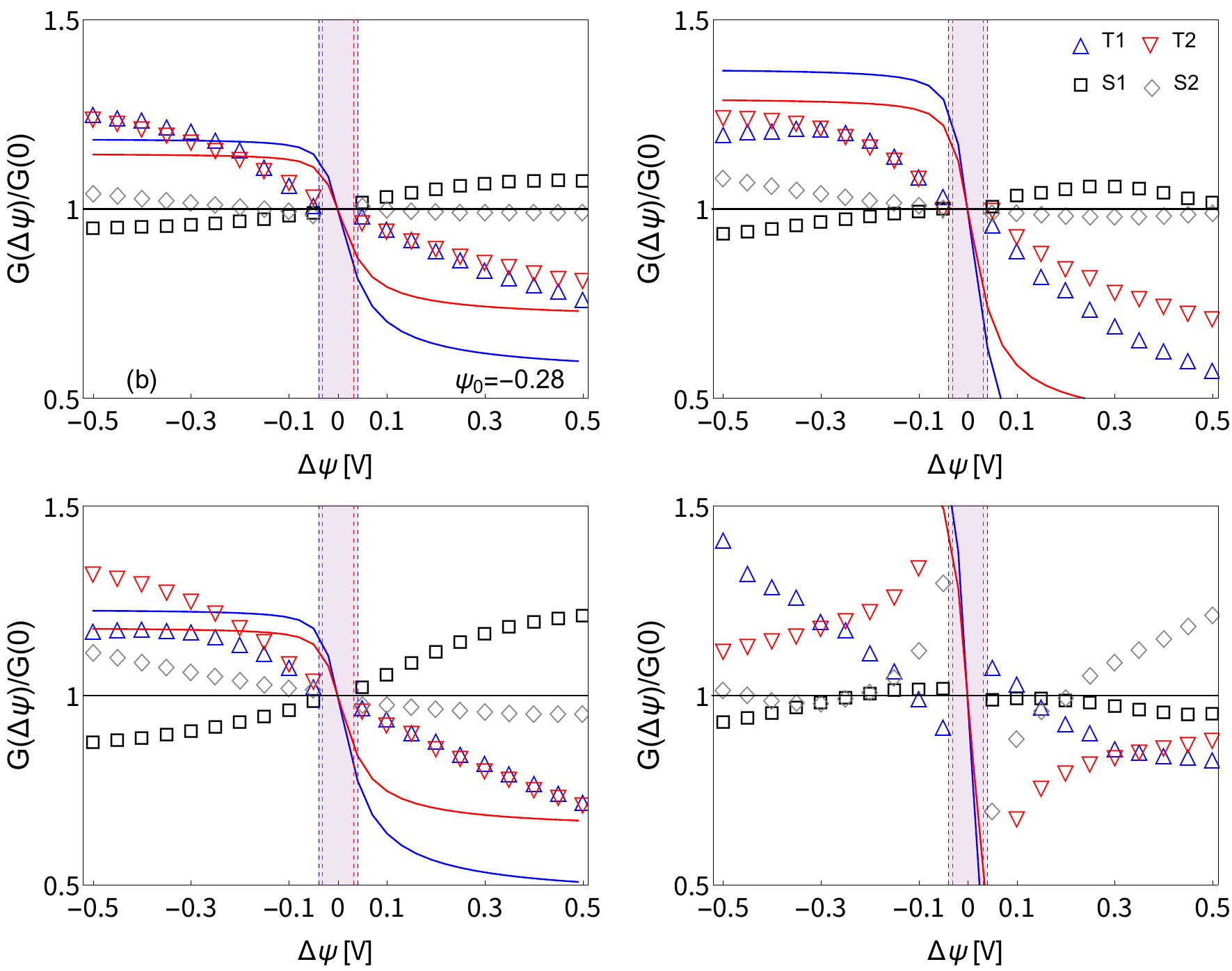}
	\caption{Reproduction of Fig.\hyref{fig:ICR}(a) in the main text at different concentrations with (a) $\rho_{\rm{b}}=1.5$ mM, (b) $\rho_{\rm{b}}=10$ mM, (c) $\rho_{\rm{b}}=6$ mM and (d) $\rho_{\rm{b}}=0.3$ mM. Figure (a) represents the best agreement between theory and experiment we could obtain, (b) and (c) are typical for our experimental results while (d) shows the large experimental variation typical at low concentrations.}
	\label{allcon}
\end{figure}

\newpage

\subsection{Supplementary Information 8: More fits of experimental data}
\begin{figure}[!htb]
	\includegraphics[width=0.4\textwidth]{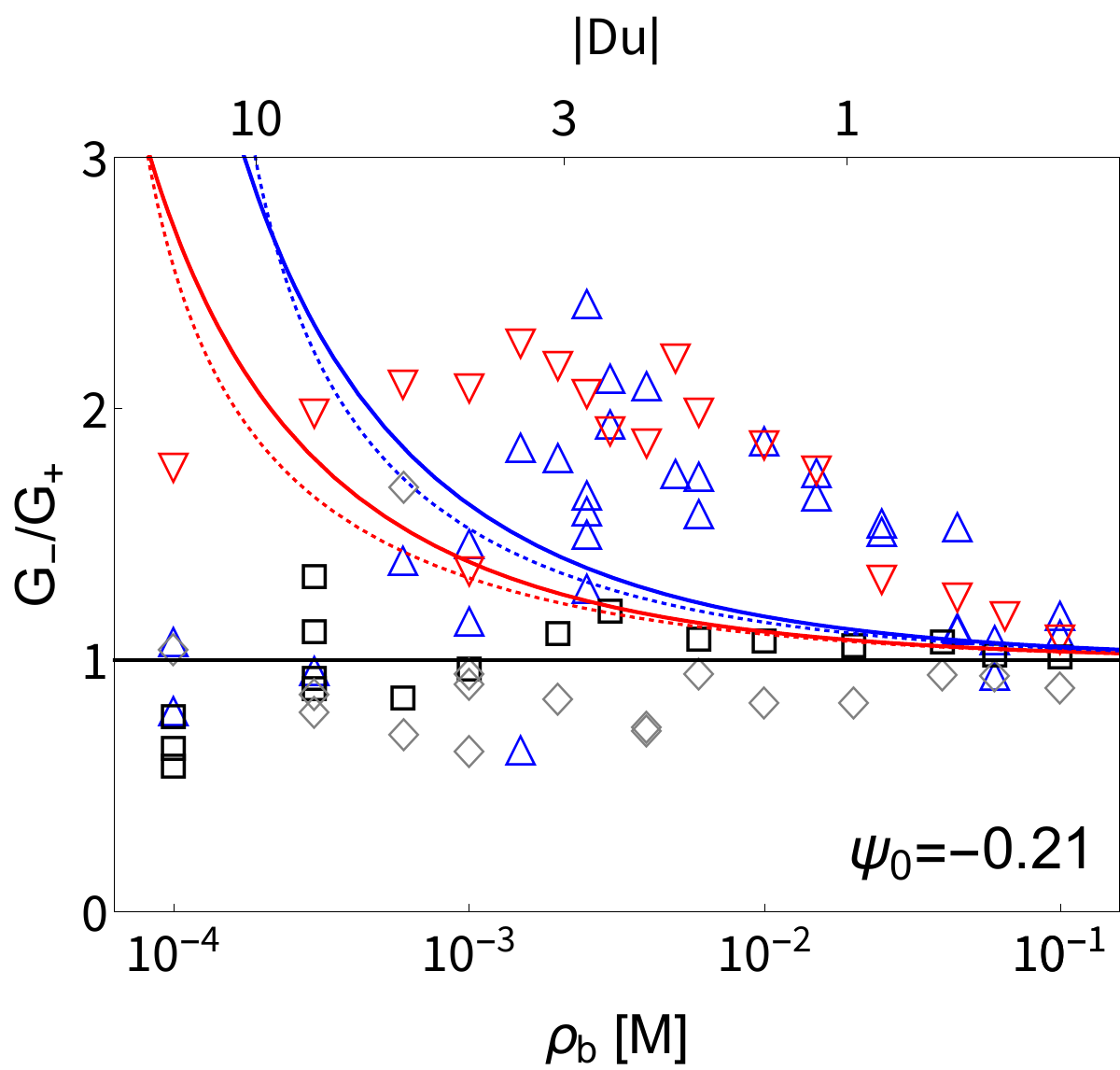}
	\caption{Current rectification $G_-/G_+$ with the surface potential $\psi_0=-0.21$ V obtained from the fit on Ohmic conductance instead of the ideal fitted surface potential for ICR $\psi_0=-0.28$ V. The quality of the fit has decreased significantly compared to Fig.(\hyref{fig:ICR}) in the main-text, but the same qualitative trend can be observed.}
	\label{ICR026}
\end{figure}
\begin{figure}[!htb]
	\includegraphics[width=0.4\textwidth]{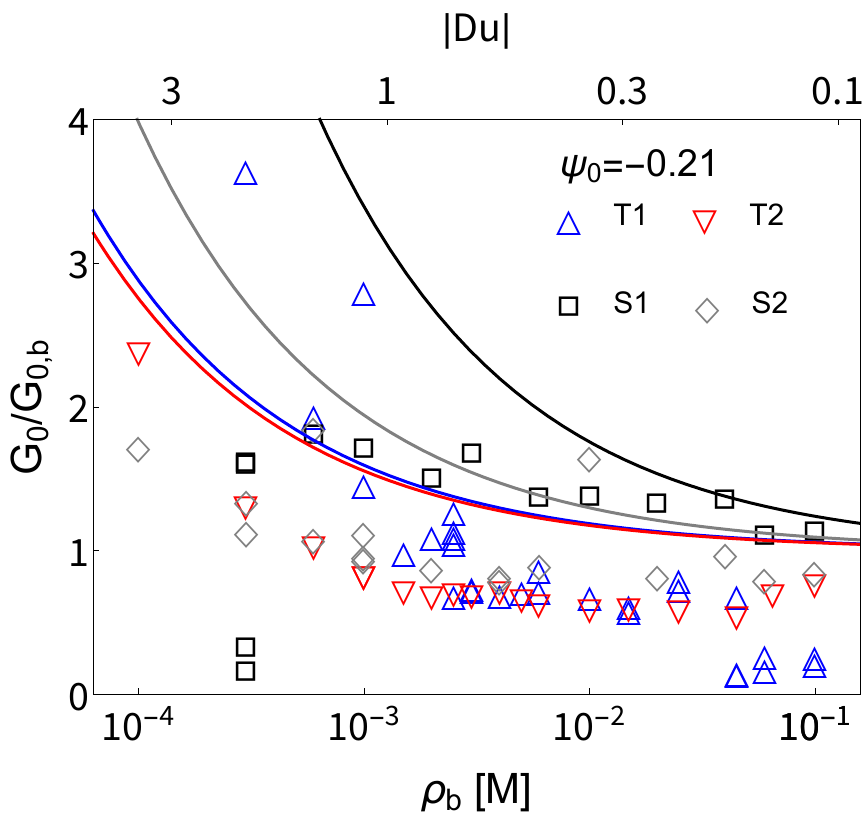}
	\caption{Fit of Ohmic conductivity $G_0$ in units of the bulk conductance $G_{0,\rm{b}}$ with a diffusion constant $D = 1.5$ nm$^{-2}$ ns$^{-1}$ instead of $D=1$nm$^{-2}$ ns$^{-1}$ as presented in Fig.3 of the main text. }
	\label{bulkcondfit}
\end{figure}

\newpage

\subsection{Supplementary Information 9: Selectivity from literature}

\begin{figure}[!htb]
	\includegraphics[width=0.6\textwidth]{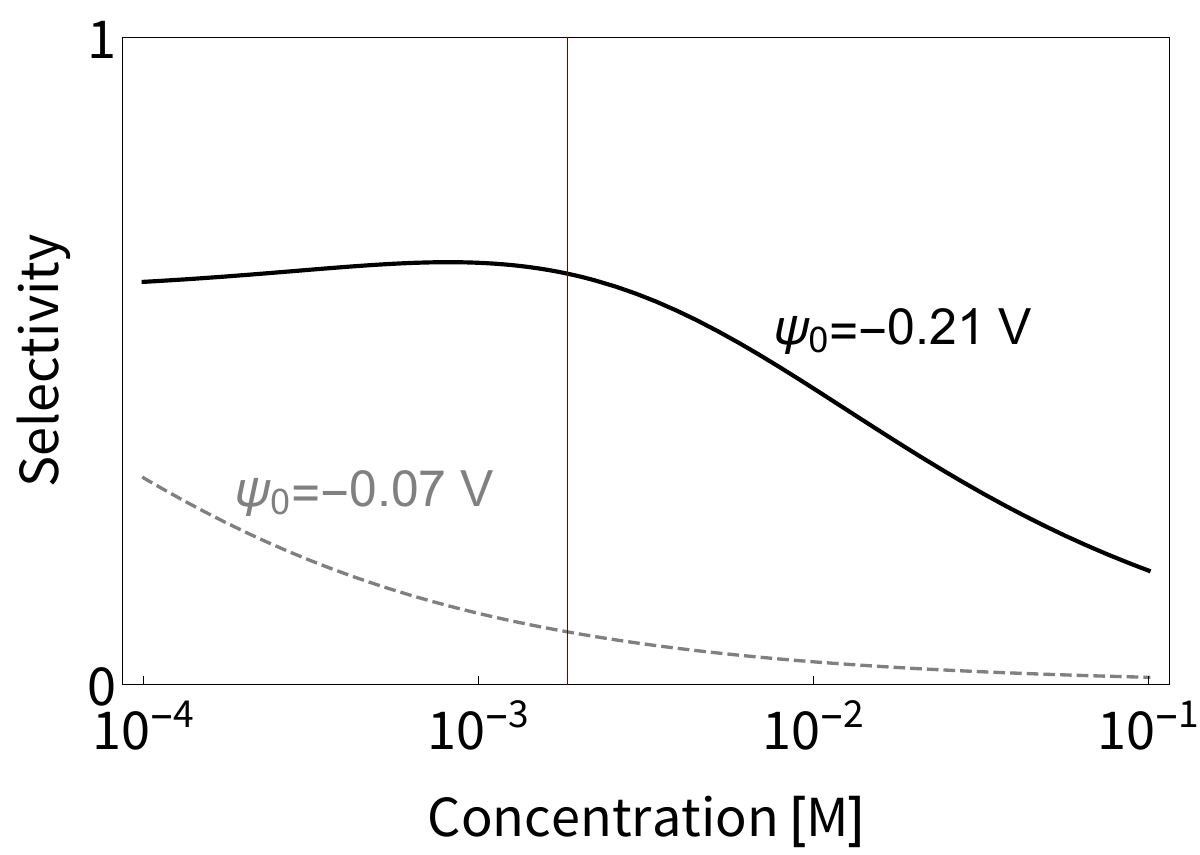}
	\caption{Selectivity as defined by Ref.\cite{Cengio2019} for our experimental geometry T1 with a surface potential $\psi_0=-0.21$ V (solid) and  $\psi_0=-0.07$ V (dashed). The selectivity shows a maximum at $\rho_{\rm{b}}\simeq 2$ mM (vertical line) for $\psi_0=-0.21$ V in line with our own experimental and theoretical results. Using a literature surface potential of $\psi_0=-0.07$ V little selectivity is expected. This shows that other theories also require an excessively high surface potential to explain the observed ICR.}
	\label{Cengio}
\end{figure}

\end{document}